\documentclass[12pt]{article}

\usepackage{axodraw}
\usepackage{feynarts}
\usepackage{graphicx}

\makeatletter

\def\paragraph{\@startsection{paragraph}{4}{\z@}{+2.00ex plus
 +1ex minus +.2ex}{1.5ex plus .2ex}{\it\normalsize}}

\def\section{\@startsection {section}{1}{\z@}{+3.0ex plus +1ex minus
  +.2ex}{2.3ex plus .2ex}{\normalsize\bf\boldmath}}
\def\subsection{\@startsection{subsection}{2}{\z@}{+2.5ex plus +1ex
minus +.2ex}{1.5ex plus .2ex}{\normalsize\bf\boldmath}}
\def\subsubsection{\@startsection{subsubsection}{3}{\z@}{+3.25ex plus
 +1ex minus +.2ex}{1.5ex plus .2ex}{\normalsize\it}}

 \expandafter\ifx\csname mathrm\endcsname\relax\def\mathrm#1{{\rm #1}}\fi

\newcounter{saveeqn}

\@addtoreset{equation}{section}

\newcount\@tempcntc
\def\@citex[#1]#2{\if@filesw\immediate\write\@auxout{\string\citation{#2}}\fi
  \@tempcnta\z@\@tempcntb\m@ne\def\@citea{}\@cite{\@for\@citeb:=#2\do
    {\@ifundefined
       {b@\@citeb}{\@citeo\@tempcntb\m@ne\@citea
        \def\@citea{,\penalty\@m\ }{\bf ?}\@warning
       {Citation `\@citeb' on page \thepage \space undefined}}%
    {\setbox\z@\hbox{\global\@tempcntc0\csname
b@\@citeb\endcsname\relax}%
     \ifnum\@tempcntc=\z@ \@citeo\@tempcntb\m@ne
       \@citea\def\@citea{,\penalty\@m}
       \hbox{\csname b@\@citeb\endcsname}%
     \else
      \advance\@tempcntb\@ne
      \ifnum\@tempcntb=\@tempcntc
      \else\advance\@tempcntb\m@ne\@citeo
      \@tempcnta\@tempcntc\@tempcntb\@tempcntc\fi\fi}}\@citeo}{#1}}

\def\@citeo{\ifnum\@tempcnta>\@tempcntb\else\@citea
  \def\@citea{,\penalty\@m}%
  \ifnum\@tempcnta=\@tempcntb\the\@tempcnta\else
   {\advance\@tempcnta\@ne\ifnum\@tempcnta=\@tempcntb \else
\def\@citea{--}\fi
    \advance\@tempcnta\m@ne\the\@tempcnta\@citea\the\@tempcntb}\fi\fi}

\def\nl{\nonumber\\}

\newcommand{\gsim}
{\mathrel{\raisebox{-.3em}{$\stackrel{\displaystyle >}{\sim}$}}}
\def\asymp#1%
{\mathrel{\raisebox{-.4em}{$\widetilde{\scriptstyle #1}$}}}

\def\Nequal#1%
{\mathrel{\raisebox{-.5em}{$\stackrel{=}{\scriptstyle\rm#1}$}}}
\newcommand{\dsl}[1]{\not \hspace{-0.7mm}#1}
\def\dsl{\mathpalette\make@slash}
\def\make@slash#1#2{\setbox\z@\hbox{$#1#2$}%
  \hbox to 0pt{\hss$#1/$\hss\kern-\wd0}\box0}

\def\beq{\begin{equation}}
\def\eeq{\end{equation}}
\def\beqar{\begin{eqnarray}}
\def\eeqar{\end{eqnarray}}
\def\barr#1{\begin{array}{#1}}
\def\earr{\end{array}}
\def\bfi{\begin{figure}}
\def\efi{\end{figure}}
\def\btab{\begin{table}}
\def\etab{\end{table}}
\def\bce{\begin{center}}
\def\ece{\end{center}}
\def\nn{\nonumber}
\def\disp{\displaystyle}
\def\text{\textstyle}

\def\al{\alpha}

\def\Ga{\Gamma}
\def\ga{\gamma}
\def\de{\delta}
\def\De{\Delta}

\def\la{\lambda}

\def\si{\sigma}

\def\refeq#1{\mbox{(\ref{#1})}}

\def\reffi#1{\mbox{Figure~\ref{#1}}}
\def\reffis#1{\mbox{Figures~\ref{#1}}}
\def\refta#1{\mbox{Table~\ref{#1}}}

\def\refse#1{\mbox{Section~\ref{#1}}}

\def\citere#1{\mbox{Ref.~\cite{#1}}}
\def\citeres#1{\mbox{Refs.~\cite{#1}}}

\newcommand{\GeV}{\unskip\,\mathrm{GeV}}
\newcommand{\MeV}{\unskip\,\mathrm{MeV}}

\newcommand{\fb}{\unskip\,\mathrm{fb}}

\newcommand{\rd}{{\mathrm{d}}}

\newcommand{\rT}{{\mathrm{T}}}

\newcommand{\Ord}{\mathswitch{{\cal{O}}}}
\newcommand{\Oa}{\mathswitch{{\cal{O}}(\alpha)}}

\newcommand{\A}{{\cal{A}}}

\newcommand{\M}{{\cal{M}}}
\def\mathswitchr#1{\relax\ifmmode{\mathrm{#1}}\else$\mathrm{#1}$\fi}

\newcommand{\PW}{\mathswitchr W}

\newcommand{\PZ}{\mathswitchr Z}

\newcommand{\Pg}{\mathswitchr g}

\newcommand{\PH}{\mathswitchr H}
\newcommand{\Pe}{\mathswitchr e}

\newcommand{\Pd}{\mathswitchr d}

\newcommand{\Pu}{\mathswitchr u}

\newcommand{\Ps}{\mathswitchr s}

\newcommand{\Pc}{\mathswitchr c}

\newcommand{\Pb}{\mathswitchr b}

\newcommand{\Pt}{\mathswitchr t}

\newcommand{\Pep}{\mathswitchr {e^+}}
\newcommand{\Pem}{\mathswitchr {e^-}}

\newcommand{\Pp}{\mathswitchr {p}}

\def\mathswitch#1{\relax\ifmmode#1\else$#1$\fi}

\newcommand{\MW}{\mathswitch {M_\PW}}

\newcommand{\MZ}{\mathswitch {M_\PZ}}
\newcommand{\MH}{\mathswitch {M_\PH}}
\newcommand{\Me}{\mathswitch {m_\Pe}}
\newcommand{\Mmy}{\mathswitch {m_\mu}}
\newcommand{\Mta}{\mathswitch {m_\tau}}
\newcommand{\Md}{\mathswitch {m_\Pd}}
\newcommand{\Mu}{\mathswitch {m_\Pu}}
\newcommand{\Ms}{\mathswitch {m_\Ps}}
\newcommand{\Mc}{\mathswitch {m_\Pc}}
\newcommand{\Mb}{\mathswitch {m_\Pb}}
\newcommand{\Mt}{\mathswitch {m_\Pt}}
\newcommand{\GW}{\Gamma_{\PW}}

\newcommand{\GZ}{\Gamma_{\PZ}}

\newcommand{\GF}{\mathswitch {G_\mu}}

\newcommand{\als}{\al_{\mathrm{s}}}

\def\solid{\raise.9mm\hbox{\protect\rule{1.1cm}{.2mm}}}
\def\dash{\raise.9mm\hbox{\protect\rule{2mm}{.2mm}}\hspace*{1mm}}

\def\ie{i.e.\ }
\def\eg{e.g.\ }

\newcommand{\LO}{{\mathrm{LO}}}
\newcommand{\NLO}{{\mathrm{NLO}}}

\newcommand{\EW}{{\mathrm{EW}}}
\newcommand{\QCD}{{\mathrm{QCD}}}
\newcommand{\QED}{{\mathrm{QED}}}
\newcommand{\LEP}{{\mathrm{LEP}}}

\newcommand{\best}{{\mathrm{best}}}
\newcommand{\tuned}{{\mathrm{tuned}}}
\newcommand{\pole}{{\mathrm{pole}}}

\newcommand{\virt}{{\mathrm{virt}}}
\newcommand{\real}{{\mathrm{real}}}
\newcommand{\fact}{{\mathrm{fact}}}

\newcommand{\MSbar}{\overline{\mathrm{MS}}}
\newcommand{\DIS}{{\mathrm{DIS}}}

\def\Re{\mathop{\mathrm{Re}}\nolimits}

\def\lra{\mathop{\mathrm{\leftrightarrow}}\nolimits}

\newcommand{\ppjjh}{\Pp\Pp\to\PH+2\mathrm{jets}+X}
\newcommand{\rj}{\mathrm{j}}
\newcommand{\muF}{\mu_{\mathrm{F}}}
\newcommand{\muR}{\mu_{\mathrm{R}}}
\newcommand{\xiF}{\xi_{\mathrm{F}}}
\newcommand{\xiR}{\xi_{\mathrm{R}}}

\newcommand{\gammainduced}{\mbox{\scriptsize $q\ga$}}
\newcommand{\gsplit}{\mbox{\scriptsize g-split}}
\newcommand{\ggfus}{\mbox{\scriptsize gg-fusion}}
\newcommand{\hhtwoloops}{\GF^2\MH^4}

\newcommand{\bqin}{\mbox{\scriptsize b-in}}
\newcommand{\bqout}{\mbox{\scriptsize b-out}}
\newcommand{\bqall}{\mbox{\scriptsize b-in/out}}

\def\draftdate{\relax}
\def\mda{\relax}
\def\mua{\relax}
\def\mla{\relax}
\def\Mda{\relax}
\def\Mua{\relax}
\def\Mla{\relax}
\def\draft{
\def\thtystars{******************************}
\def\sixtystars{\thtystars\thtystars}
\typeout{}
\typeout{\sixtystars**}
\typeout{* Draft mode!
         For final version remove \protect\draft\space in source file *}
\typeout{\sixtystars**}
\typeout{}
\def\draftdate{\today}
\def\mua{\marginpar[\boldmath\hfil$\uparrow$]%
                   {\boldmath$\uparrow$\hfil}%
                    \typeout{marginpar: $\uparrow$}\ignorespaces}
\def\mda{\marginpar[\boldmath\hfil$\downarrow$]%
                   {\boldmath$\downarrow$\hfil}%
                    \typeout{marginpar: $\downarrow$}\ignorespaces}
\def\mla{\marginpar[\boldmath\hfil$\rightarrow$]%
                   {\boldmath$\leftarrow $\hfil}%
                    \typeout{marginpar: $\lra$}\ignorespaces}
\def\Mua{\marginpar[\boldmath\hfil$\Uparrow$]%
                   {\boldmath$\Uparrow$\hfil}%
                    \typeout{marginpar: $\uparrow$}\ignorespaces}
\def\Mda{\marginpar[\boldmath\hfil$\Downarrow$]%
                   {\boldmath$\Downarrow$\hfil}%
                    \typeout{marginpar: $\downarrow$}\ignorespaces}
\def\Mla{\marginpar[\boldmath\hfil$\Rightarrow$]%
                   {\boldmath$\Leftarrow $\hfil}%
                    \typeout{marginpar: $\lra$}\ignorespaces}
\overfullrule 5pt
\oddsidemargin -15mm
\marginparwidth 29mm
}

\def\stars{\strut\leaders\hbox{*}\hfill\strut}
\def\starline{\hfil\strut\hfil\hbox to \textwidth {\stars}\hfil}

\begin{document}
\thispagestyle{empty}
\def\thefootnote{\fnsymbol{footnote}}
\setcounter{footnote}{1}
\null
\draftdate\hfill MPP-2007-152\\
\strut\hfill PSI-PR-07-06\\
\strut\hfill UWThPh-2007-26\\
%\strut\hfill HEPTOOLS 07-020\\
\vspace{1.5cm}
\begin{center}
{\Large \bf\boldmath
Electroweak and QCD corrections to Higgs production via
vector-boson fusion at the LHC
\\[.5em]
\par} 
\vspace{1cm}
{\large
{\sc M.\ Ciccolini$^1$, A.\ Denner$^1$ and S.\ Dittmaier$^{2,3}$} 
} \\[1cm]
$^1$ {\it Paul Scherrer Institut, W\"urenlingen und Villigen,
\\
CH-5232 Villigen PSI, Switzerland} \\[0.5cm]
$^2$ {\it Max-Planck-Institut f\"ur Physik
(Werner-Heisenberg-Institut), \\
%F\"ohringer Ring 6,
D-80805 M\"unchen, Germany}\\[0.5cm]
$^3$ {\it Faculty of Physics, University of Vienna, \\
A-1090 Vienna, Austria}
\\[0.5cm]
\par \vskip 2em
\end{center}\par
\vfill {\bf Abstract:} \par The radiative corrections of the strong
and electroweak interactions are calculated at next-to-leading order
for Higgs-boson production in the weak-boson-fusion channel at hadron
colliders.  Specifically, the calculation includes all weak-boson
fusion and quark--antiquark annihilation diagrams to Higgs-boson
production in association with two hard jets, including all
corresponding interferences.  The results on the QCD corrections
confirm that previously made approximations of neglecting $s$-channel
diagrams and interferences are well suited for predictions of Higgs
production with dedicated vector-boson fusion cuts at the LHC. The
electroweak corrections, which also include real corrections from
incoming photons and leading heavy-Higgs-boson effects at two-loop
order, are of the same size as the QCD corrections, viz.\ typically at
the level of $5{-}10\%$ for a Higgs-boson mass up to $\sim700\GeV$.
In general, both types of corrections do not simply rescale
differential distributions, but induce distortions at the level of
10\%.  The discussed corrections have been implemented in a flexible
Monte Carlo event generator.
\par
\vskip .5cm
\noindent
October 2007 
\null
\setcounter{page}{0}
\clearpage
\def\thefootnote{\arabic{footnote}}
\setcounter{footnote}{0}

\section{Introduction}

The production of a Standard Model Higgs boson in association with two
hard jets in the forward and backward regions of the
detector---frequently quoted as the ``vector-boson fusion'' (VBF)
channel---is a cornerstone in the Higgs search both in the ATLAS
\cite{Asai:2004ws} and CMS \cite{Abdullin:2005yn} experiments at the
LHC. This is not only true for the Higgs-mass range between 100 and
$200\GeV$, which is favoured by the global Standard Model fit to
electroweak (EW) precision data \cite{Alcaraz:2006mx}, but also for a
Higgs mass of the order of several $100\GeV$ up to the theoretical
upper limit set by unitarity and triviality constraints.  Higgs
production in the VBF channel also plays an important role in the
determination of Higgs couplings at the LHC (see e.g.
\citere{Duhrssen:2004cv}). Even bounds on non-standard couplings
between Higgs and EW gauge bosons can be imposed from precision
studies in this channel \cite{Hankele:2006ma}.

The production of Higgs+2jets receives two kinds of contributions at
hadron colliders. The first type, where the Higgs boson couples to a
weak boson that links two quark lines, is dominated by squared $t$-
and $u$-channel-like diagrams and represents the genuine VBF channel.
The hard jet pairs have a strong tendency to be forward--backward
directed in contrast to other jet production mechanisms, offering a
good background suppression (transverse-momentum and rapidity cuts on
jets, jet rapidity gap, central-jet veto, etc.).  Applying appropriate
event selection criteria (see
e.g.~\citeres{Barger:1994zq,Rainwater:1997dg,Rainwater:1998kj,%
  Rainwater:1999sd,DelDuca:2006hk} and references in
\citeres{Spira:1997dg,Djouadi:2005gi}) it is possible to sufficiently
suppress background and to enhance the VBF channel over the second
Higgs+2jets production mechanism that mainly proceeds via strong
interactions. In this second channel the Higgs boson is radiated off a
heavy-quark loop that couples to any parton of the incoming hadrons
via gluons \cite{DelDuca:2001fn,Campbell:2006xx}.  According to a
recent estimate \cite{Nikitenko:2007it} hadronic production
contributes about $4{-}5\%$ to the Higgs+2jets events for a Higgs mass
of $120\GeV$ after applying VBF cuts. A next-to-leading order (NLO)
analysis of this contribution \cite{Campbell:2006xx} shows that its
residual scale dependence is still of the order of 35\%.

Higgs production in the VBF channel is a pure EW process in leading
order (LO) involving only quark and antiquark parton distributions.
As $s$-channel diagrams and interferences tend to be suppressed,
especially when imposing VBF cuts, the cross section can be
approximated by the contribution of squared $t$- and $u$-channel
diagrams only.  The corresponding QCD corrections reduce to vertex
corrections to the weak-boson--quark coupling.  Explicit NLO QCD
calculations in this approximation
\cite{Spira:1997dg,Han:1992hr,Figy:2003nv,Figy:2004pt,Berger:2004pc}
confirm the expectation that these QCD corrections are small, because
they are shifted to the parton distribution functions (PDFs) via QCD
factorization to a large extent. The resulting QCD corrections are of
the order of $5{-}10\%$ and reduce the remaining factorization and
renormalization scale dependence of the NLO cross section to a few per
cent.

In a recent letter \cite{Ciccolini:2007jr} we completed the existing
NLO calculations for the VBF channel in two respects. Firstly, we
added the full NLO EW corrections. Secondly, we calculated the NLO QCD
corrections including, for the first time, the complete set of QCD
diagrams, namely the $t$-, $u$-, and $s$-channel contributions, as
well as all interferences.  Focussing on the integrated cross section
(with and without dedicated VBF selection cuts), we discussed the
impact of EW and QCD corrections in the favoured Higgs-mass range
between 100 and $200\GeV$.  We found that the previously unknown NLO
EW corrections are of the order of $-5\%$ and, thus, as important as
the QCD corrections. In the EW corrections we also take into account
real corrections induced by photons in the initial state and QED
corrections implicitly contained in the DGLAP evolution of PDFs. We
found that these photon-induced processes lead to corrections at the
per-cent level.

In this paper we describe more details of our calculation, which is
performed in a widely analogous way to the EW and QCD corrections to
the Higgs decay $\PH\to \PW\PW/\PZ\PZ\to4\,$fermions
\cite{Bredenstein:2006rh,Bredenstein:2006ha}.  We classify the NLO QCD
corrections into four different categories; the previously known
corrections
\cite{Spira:1997dg,Han:1992hr,Figy:2003nv,Figy:2004pt,Berger:2004pc}
are contained in one of these categories.  Moreover, we extend our
numerical discussion of the EW and QCD corrections in two respects. We
now consider cross sections for Higgs masses above $200\GeV$,
including the leading EW two-loop corrections $\propto\GF^2\MH^4$ for
a heavy Higgs boson using the results of \citere{Ghinculov:1995bz},
and we discuss differential distributions in transverse momenta, in
rapidities, in the azimuthal angle difference of the tagging jets, and
in the jet--jet invariant mass.  We pay particular attention to the
issue of distortions in distributions induced by radiative
corrections, because such distortions usually are the signature of
non-standard couplings.

The paper is organized as follows: In \refse{se:setup} we describe the
calculational setup and classify the QCD corrections into different
categories. The discussion of numerical results is presented in
\refse{se:numres}. Section~\ref{se:concl} contains our conclusions.

\section{Details of the calculation}
\label{se:setup}

\subsection{General setup}

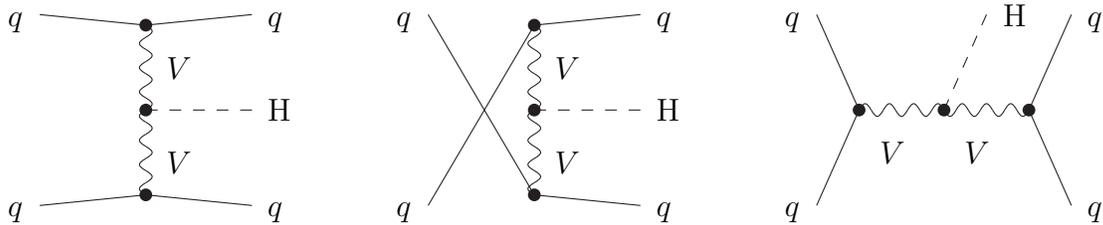
\begin{figure}
\centerline{
{\unitlength .8pt %\scriptsize
\begin{picture}(125,100)(0,0)
\SetScale{.8}
\Line( 15,95)( 65,90)
\Line( 15, 5)( 65,10)
\Line(115, 5)( 65,10)
\Line(115,95)( 65,90)
\Photon( 65,10)( 65,90){3}{7}
\DashLine(115,50)( 65,50){6}
\Vertex(65,90){3}
\Vertex(65,10){3}
\Vertex(65,50){3}
\put(  0,90){$q$}
\put(  0,0){$q$}
\put(123,90){{$q$}}
\put(123,0){{$q$}}
\put(123,45){{$\PH$}}
\put( 75,65){{$V$}}
\put( 75,20){{$V$}}
\SetScale{1}
\end{picture}
}
\hspace*{3em}
{\unitlength .8pt %\scriptsize
\begin{picture}(125,100)(0,0)
\SetScale{.8}
\Line( 15,95)( 65,10)
\Line( 15, 5)( 65,90)
\Line(115, 5)( 65,10)
\Line(115,95)( 65,90)
\Photon( 65,10)( 65,90){3}{7}
\DashLine(115,50)( 65,50){6}
\Vertex(65,90){3}
\Vertex(65,10){3}
\Vertex(65,50){3}
\put(  0,90){$q$}
\put(  0,0){$q$}
\put(123,90){{$q$}}
\put(123,0){{$q$}}
\put(123,45){{$\PH$}}
\put( 75,65){{$V$}}
\put( 75,20){{$V$}}
\SetScale{1}
\end{picture}
}
\hspace*{3em}
{\unitlength .8pt %\scriptsize
\begin{picture}(145,100)(0,0)
\SetScale{.8}
\Line( 15,95)( 35,50)
\Line( 15, 5)( 35,50)
\Line(135,95)(115,50)
\Line(135, 5)(115,50)
\Photon( 35,50)(115,50){3}{7}
\DashLine(75,50)(95,95){6}
\Vertex( 35,50){3}
\Vertex(115,50){3}
\Vertex( 75,50){3}
\put(  0,90){$q$}
\put(  0,0){$q$}
\put(143,90){{$q$}}
\put(143,0){{$q$}}
\put(103,90){{$\PH$}}
\put( 45,25){{$V$}}
\put( 85,25){{$V$}}
\SetScale{1}
\end{picture}
}
}
\caption{Topologies for $t$-, $u$-, and $s$-channel contributions to
  $qq\to qq\PH$ in LO, where $q$ denotes any quark or antiquark and
  $V$ stands for W and Z~bosons.}
\label{fig:LOtops}
\end{figure}

At LO, the hadronic production of Higgs+2jets via weak bosons receives
contributions from the partonic processes $qq\to qq\PH$, $q\bar{q}\to
q\bar{q}\PH$, $\bar{q}q\to \bar{q}q\PH$, and
$\bar{q}\bar{q}\to\bar{q}\bar{q}\PH$.  For each relevant configuration
of external quark flavours one or two of the topologies shown in
\reffi{fig:LOtops} contribute.
All LO and one-loop NLO diagrams are related by crossing symmetry to
the corresponding decay amplitude $\PH\to q\bar{q}q\bar{q}$. The QCD
and EW NLO corrections to these decays were discussed in detail in
\citeres{Bredenstein:2006rh,Bredenstein:2006ha}, in particular a
representative set of Feynman diagrams can be found there.

\begin{sloppypar}
  To be more specific, we first show how the lowest-order and loop
  amplitudes for subprocesses of the type $\bar{q}q\to\bar{q}q\PH$ can
  be obtained from the corresponding results for $\PH\to
  q_a\bar{q}_bq_c\bar{q}_d$.  The basic lowest-order decay amplitudes
  involving $\PW$- or $\PZ$-boson exchange, called
  $\M^{VV,\si_a\si_b\si_c\si_d}_0(k_a,k_b,k_c,k_d)$ with $V=\PW,\PZ$,
  have been defined in Eq.~(2.8) of \citere{Bredenstein:2006rh}; the
  two potentially relevant tree diagrams are shown in
  \reffi{fi:H4f-born-diags}; the corresponding squares and
  interference are illustrated in \reffi{fi:H4f-born-ints}.
\bfi
\centerline{
\setlength{\unitlength}{1pt}
\begin{picture}(190,100)(-20,0)
\DashLine(15,50)(60,50){3}
\Photon(60,50)(90,20){-2}{5}
\Photon(60,50)(90,80){2}{5}
\Vertex(60,50){2.0}
\Vertex(90,80){2.0}
\Vertex(90,20){2.0}
\ArrowLine(90,80)(120, 95)
\ArrowLine(120,65)(90,80)
\ArrowLine(120, 5)( 90,20)
\ArrowLine( 90,20)(120,35)
\put(0,47){$\PH$}
\put(62,70){$V$}
\put(62,18){$V$}
\put(125,90){$q_a$}
\put(125,65){$\bar q_b$}
\put(125,30){$q_c$}
\put(125,5){$\bar q_d$}
\end{picture}
\begin{picture}(190,100)(-20,0)
\DashLine(15,50)(60,50){3}
\Photon(60,50)(90,35){-2}{4}
\Photon(60,50)(90,80){2}{5}
\Vertex(60,50){2.0}
\Vertex(90,80){2.0}
\Vertex(90,35){2.0}
\ArrowLine(90,80)(120, 95)
\ArrowLine(120,65)(90,35)
\ArrowLine(120, 5)(90,80)
\ArrowLine( 90,35)(120,35)
\put(0,47){$\PH$}
\put(62,70){$V'$}
\put(62,18){$V'$}
\put(125,90){$q_a$}
\put(125,65){$\bar q_b$}
\put(125,30){$q_c$}
\put(125,5){$\bar q_d$}
\end{picture}
}
\caption{Generic lowest-order diagrams for 
  the Higgs decay $\PH\to q_a\bar q_b q_c\bar q_d$, where
  $V,V'=\PW,\PZ$ denote the exchanged weak bosons. The lowest-order
  diagrams for $qq\to qq\PH$ are obtained by crossing 
  any pair of (anti-)quarks into the initial state and the Higgs boson
  into the final state.}
\label{fi:H4f-born-diags}
\vspace*{3em}
\centerline{
\setlength{\unitlength}{1pt}
\begin{picture}(180,100)
\DashLine( 0,50)(30,50){3}
\Photon(30,50)(60,20){-2}{5}
\Photon(30,50)(60,80){2}{5}
\Vertex(30,50){2.0}
\Vertex(60,80){2.0}
\Vertex(60,20){2.0}
\ArrowLine(60,80)(90,95)
\ArrowLine(90,65)(60,80)
\ArrowLine(90, 5)(60,20)
\ArrowLine(60,20)(90,35)
\DashLine(90,0)(90,100){7}
\ArrowLine(90,95)(120,80)
\ArrowLine(120,80)(90,65)
\ArrowLine(120,20)(90, 5)
\ArrowLine(90,35)(120,20)
\Vertex(150,50){2.0}
\Vertex(120,80){2.0}
\Vertex(120,20){2.0}
\Photon(150,50)(120,20){2}{5}
\Photon(150,50)(120,80){-2}{5}
\DashLine(180,50)(150,50){3}
\put(0,90){(A)}
\end{picture}
\hspace*{2em}
\begin{picture}(180,100)
\DashLine(0,50)(30,50){3}
\Photon(30,50)(60,35){-2}{4}
\Photon(30,50)(60,80){2}{5}
\Vertex(30,50){2.0}
\Vertex(60,80){2.0}
\Vertex(60,35){2.0}
\ArrowLine(60,80)(90,95)
\ArrowLine(90,65)(60,35)
\ArrowLine(90, 5)(60,80)
\ArrowLine(60,35)(90,35)
\DashLine(90,0)(90,100){7}
\ArrowLine(90,95)(120,80)
\ArrowLine(120,80)(90,65)
\ArrowLine(120,20)(90, 5)
\ArrowLine(90,35)(120,20)
\Vertex(150,50){2.0}
\Vertex(120,80){2.0}
\Vertex(120,20){2.0}
\Photon(150,50)(120,20){2}{5}
\Photon(150,50)(120,80){-2}{5}
\DashLine(180,50)(150,50){3}
\put(0,90){(B)}
\end{picture}
}
\caption{Types of squared and interference diagrams contributing in
  lowest order.} 
\label{fi:H4f-born-ints}
\efi
The external momenta $\{p_i,p'_i\}$, helicities $\{\si_i,\si'_i\}$,
and colour indices $\{c_i,c'_i\}$ are assigned to the scattering
particles according to
\beq\label{eq:qbarqqbarq}
\bar{q}(p_1,\si_1,c_1) + q(p_2,\si_2,c_2) \to 
\bar{q}(p'_1,\si'_1,c'_1) + q(p'_2,\si'_2,c'_2) + \PH(p'_3).
\eeq
In order to compactify notation, we omit the labels $p_i$, $\si_i$,
$c_i$, etc.  in the amplitudes $\A^{\bar qq\to\bar qq}$ of the
scattering process, i.e.\ we implicitly have $\A^{\bar qq\to\bar qq}
\equiv \A^{\bar qq\to\bar qq,\si_1\si_2\si'_1\si'_2}_{c_1c_2c'_1c'_2}
(p_1,p_2,p'_1,p'_2)$, and abbreviate the helicity and momentum
assignment in $\M^{VV}$ as $\M^{VV}(122'1') \equiv
\M^{VV,-\si_1,-\si_2,\si'_2,\si'_1}(-p_1,-p_2,p'_2,p'_1)$, etc..  Note
that momenta and helicities crossed into the initial state receive a
sign change. In this notation the lowest-order amplitudes $\A_0$ for
the six basic flavour channels in $\bar{q}q\to\bar{q}q\PH$ read
\beqar
\A_0^{\bar u_i d_j\to \bar u_k d_l}
&=& V_{ij} V_{kl}^* C_1^{c_1c_2c'_2c'_1} \M^{\PW\PW}_0(122'1')
- \de_{ik}\de_{jl}  C_1^{c_1c'_1c'_2c_2} \M^{\PZ\PZ}_0(11'2'2),
\nn\\
\A_0^{\bar d_i u_j\to \bar d_k u_l}
&=& V_{ji}^* V_{lk} C_1^{c'_2c'_1c_1c_2} \M^{\PW\PW}_0(2'1'12)
- \de_{ik}\de_{jl}  C_1^{c'_2c_2c_1c'_1} \M^{\PZ\PZ}_0(2'211'),
\nn\\
\A_0^{\bar u_i u_j\to \bar d_k d_l}
&=& \de_{ij}\de_{kl} C_1^{c_1c_2c'_2c'_1} \M^{\PZ\PZ}_0(122'1')
- V_{ik} V_{jl}^*    C_1^{c_1c'_1c'_2c_2} \M^{\PW\PW}_0(11'2'2),
\nn\\
\A_0^{\bar d_i d_j\to \bar u_k u_l}
&=& \de_{ij}\de_{kl} C_1^{c'_2c'_1c_1c_2} \M^{\PZ\PZ}_0(2'1'12)
- V_{ki}^* V_{lj}    C_1^{c'_2c_2c_1c'_1} \M^{\PW\PW}_0(2'211'),
\nn\\
\A_0^{\bar u_i u_j\to \bar u_k u_l}
&=& \de_{ij}\de_{kl} C_1^{c_1c_2c'_2c'_1} \M^{\PZ\PZ}_0(122'1')
- \de_{ik}\de_{jl}   C_1^{c_1c'_1c'_2c_2} \M^{\PZ\PZ}_0(11'2'2),
\nn\\
\A_0^{\bar d_i d_j\to \bar d_k d_l}
&=& \de_{ij}\de_{kl} C_1^{c_1c_2c'_2c'_1} \M^{\PZ\PZ}_0(122'1')
- \de_{ik}\de_{jl}   C_1^{c_1c'_1c'_2c_2} \M^{\PZ\PZ}_0(11'2'2),
\label{eq:Mqbarqqbarq}
\eeqar
where $i,j,k,l$ are generation indices, $V_{ij}$ are quark-mixing matrix
elements, and $C_1^{abcd}$ is one of the two colour operators
\beq
C_1^{abcd}=\de_{ab}\otimes\de_{cd}, \qquad
C_2^{abcd}=\frac{1}{4C_{\mathrm{F}}}\sum_h\la^h_{ab}\otimes\la^h_{cd} =
\frac{3}{16}\sum_h\la^h_{ab}\otimes\la^h_{cd},
\label{eq:Ccolour}
\eeq
which are relevant to span a general $\PH\to q_a\bar{q}_bq_c\bar{q}_d$
amplitude in colour space. The second operator $C_2^{abcd}$, which
involves the Gell-Mann matrices $\la^h$, becomes relevant in the QCD
corrections discussed below.  The relative sign between the two
amplitude contributions on the r.h.s.\ of \refeq{eq:Mqbarqqbarq}
originates from their different fermion-number flow.  In Section~2 of
\citere{Bredenstein:2006rh} the calculation of $\M^{VV}_0$ is
described in terms of Weyl--van-der-Waerden spinor products $\langle
ab\rangle$ in the conventions of \citere{Dittmaier:1998nn}, where $a$
and $b$ are spinors corresponding to external momenta. We note that
complex conjugate products $\langle ab\rangle^*$ (but not 
$\langle ab\rangle$) receive an additional sign factor for each crossed
momentum $-p_i$ involved in the product.
\end{sloppypar}

The lowest-order and loop amplitudes for subprocesses of the type
$qq\to {q}{q}\PH$ and $\bar q\bar q\to\bar{q}\bar{q}\PH$ can be
obtained as follows. We assign the external momenta, helicities, and
colour indices as
\beqar
{q}(p_1,\si_1,c_1) + q(p_2,\si_2,c_2) &\to& 
{q}(p'_1,\si'_1,c'_1) + {q}(p'_2,\si'_2,c'_2) + \PH(p'_3),\nl
\bar{q}(p_1,\si_1,c_1) + \bar{q}(p_2,\si_2,c_2) &\to& 
\bar{q}(p'_1,\si'_1,c'_1) + \bar{q}(p'_2,\si'_2,c'_2) + \PH(p'_3).
\eeqar
Then the corresponding amplitudes can be obtained via crossing
symmetry from those for the process \refeq{eq:qbarqqbarq} as
\beqar
\A^{qq\to qq,\si_1\si_2\si'_1\si'_2}_{c_1c_2c'_1c'_2}
(p_1,p_2,p'_1,p'_2)
&=&
\A^{\bar qq\to\bar qq,-\si'_1,\si_2,-\si_1,\si'_2}_{c'_1c_2c_1c'_2}
(-p'_1,p_2,-p_1,p'_2),\nl
\A^{\bar q\bar q\to \bar q \bar q,\si_1\si_2\si'_1\si'_2}_{c_1c_2c'_1c'_2}
(p_1,p_2,p'_1,p'_2)
&=&
\A^{\bar qq\to\bar qq,\si_1,-\si'_2,\si'_1,-\si_2}_{c_1c'_2c'_1c_2}
(p_1,-p'_2,p'_1,-p_2).
\eeqar
When calculating the corresponding cross sections, symmetry factors
1/2 must be taken into account for identical fermions or antifermions
in the final state.

In our calculation we neglect external quark masses whenever possible,
i.e.\ everywhere but in the mass-singular logarithms. In
\refeq{eq:Mqbarqqbarq} we made the CKM matrix elements $V_{ij}$
explicit.  Note that only absolute values of the CKM matrix elements
survive after squaring the amplitudes; for the squared W-mediated
diagrams this is obvious, for the interference between W- and
Z-mediated diagrams $|V_{ij}|^2$ results after contraction of the CKM
matrix elements with Kronecker deltas.  Numerically, only the mixing
among the first two generations could be relevant, but its impact on
Higgs production via VBF was found to be negligible.  Since the
contributions of external b~quarks are suppressed, either by bottom
densities or by $s$-channel suppression, we optionally include
b~quarks in the initial and final states in our LO predictions, but
not in the calculation of corrections.

\subsection{Evaluation of NLO corrections}
\label{se:nlocorr}

Evaluating $2\to3$ particle processes at the NLO level is non-trivial,
both in the analytical and numerical parts of the calculation.  In
order to ensure the correctness of our results we have evaluated each
ingredient twice, resulting in two completely independent computer
codes yielding results in mutual agreement.  The actual calculation of
virtual and real NLO corrections for the partonic processes is
performed along the same lines as described in
\citere{Bredenstein:2006rh,Bredenstein:2006ha} for the decays
$\PH\to4f$.  Therefore, we only repeat the salient features of the
evaluation.

\paragraph{Virtual corrections}

The virtual corrections modify the partonic processes that are already
present at LO; there are about 200 one-loop diagrams per tree diagram
in each flavour channel.  At NLO these corrections are induced by
self-energy, vertex, box (4-point), and pentagon (5-point) diagrams.
The calculation of the EW one-loop diagrams has been performed both in
the conventional 't~Hooft--Feynman gauge and in the background-field
formalism using the conventions of \citeres{Denner:1991kt} and
\cite{Denner:1994xt}, respectively. The QCD one-loop diagrams are
evaluated in 't~Hooft--Feynman gauge.

In contrast to the $t$- and $u$-channel contributions (first two
diagrams in \reffi{fig:LOtops}), the $s$-channel diagrams (last
diagram in \reffi{fig:LOtops}) contain resonant W- or Z-boson
propagators that require a proper inclusion of the finite gauge-boson
widths.  For the implementation of the finite widths we use the
complex-mass scheme, which was introduced in \citere{Denner:1999gp}
for lowest-order calculations and generalized to the one-loop level in
\citere{Denner:2005fg}. In this approach the W- and Z-boson masses are
consistently considered as complex quantities, defined as the
locations of the propagator poles in the complex plane.  This leads to
complex couplings and, in particular, a complex weak mixing angle.
The scheme fully respects all relations that follow from gauge
invariance.  A brief description of this scheme can also be found in
\citere{Denner:2006ic}.

The amplitudes have been generated with {\sc FeynArts}, using the two
independent versions 1 and 3, as described in
\citeres{Kublbeck:1990xc} and \cite{Hahn:2000kx}, respectively.  The
algebraic evaluation has been performed in two completely independent
ways. One calculation is based on an in-house program written in {\sl
  Mathematica}, the other has been completed with the help of {\sc
  FormCalc} \cite{Hahn:1998yk}. The amplitudes are expressed in terms
of standard matrix elements and coefficients, which contain the tensor
integrals, as described in the appendix of \citere{Denner:2003iy}.

The tensor integrals are evaluated as in the calculation of the
corrections to $\Pep\Pem\to4f$ \cite{Denner:2005fg,Denner:2005es}.
They are recursively reduced to master integrals at the numerical
level.  The scalar master integrals are evaluated for complex masses
using the methods and results of \citere{'tHooft:1979xw}.  UV
divergences are regulated dimensionally and IR divergences with an
infinitesimal photon or gluon mass.  Tensor and scalar 5-point
functions are directly expressed in terms of 4-point integrals
\cite{Denner:2002ii,Denner:2005nn}.  Tensor 4-point and 3-point
integrals are reduced to scalar integrals with the Passarino--Veltman
algorithm \cite{Passarino:1979jh} as long as no small Gram determinant
appears in the reduction. If small Gram determinants occur, we expand
the tensor coefficients about the limit of vanishing Gram determinants
and possibly other kinematical determinants, as described in
\citere{Denner:2005nn} in detail.

Since corrections due to Higgs-boson self-interactions become
important for large Higgs-boson masses, we have included the dominant
two-loop corrections to the $VV\PH$ vertex proportional to $G_\mu^2
\MH^4$ in the large-Higgs-mass limit which were calculated in
\citere{Ghinculov:1995bz}.  Specifically, we include this effect via a
correction factor
\beq
\de_{\hhtwoloops} = 
62.0308(86) \left(\frac{\GF\MH^2}{16\pi^2\sqrt{2}}\right)^2
\eeq
to the squares of the basic LO amplitudes $\M^{VV}_0$ in the $t$- and
$u$-channel. We do not include this correction in the (suppressed)
$s$-channel contributions, because the underlying assumption in the
derivation of $\de_{\GF^2\MH^4}$ that $\MH$ is much larger than any
other relevant scale is spoiled by the invariant $s$ that can be of
the order of $\MH^2$ or larger.  We do not apply $\de_{\GF^2\MH^4}$ to
interferences either, because this would require a more complicated
structure in the correction (involving more than one form factor).
The impact of ${\cal O}(\GF^2\MH^4)$ corrections on interferences and
$s$-channel contributions is certainly negligible, since these effects
are suppressed themselves.

\paragraph{Real corrections}

The matrix elements for the real corrections (photonic/gluonic
bremsstrahlung and photon-/gluon-induced processes) are obtained via
crossing from the bremsstrahlung corrections to the related Higgs
decays, $\PH\to4f+\gamma/\Pg$. Explicit amplitudes for
$\PH\to4f+\gamma$ are given in Section~4.1 of
\citere{Bredenstein:2006rh} in terms of spinor products; for
$\PH\to4f+\Pg$ such results can be found in Section~3.3 of
\citere{Bredenstein:2006ha}.  The matrix elements relevant for the
calculation presented here have been checked against results obtained
with {\sc Madgraph} \cite{Stelzer:1994ta}.

\begin{sloppypar}
  The bremsstrahlung corrections involve singularities from soft or
  collinear photon/gluon emission; the photon-/gluon-induced processes
  contain singularities from collinear initial-state splittings. Soft
  singularities, which are regularized by an infinitesimal
  photon/gluon mass, cancel between virtual and bremsstrahlung
  corrections. Collinear singularities connected to the initial or
  final state are regularized by small quark masses, which appear only
  in logarithms. While singularities connected to collinear
  configurations in the final state cancel for ``collinear-safe''
  observables automatically after applying a jet algorithm,
  singularities connected to collinear initial-state splittings are
  removed via factorization by PDF redefinitions, as described in more
  detail in \refse{se:hadxsection}.
\end{sloppypar}

Technically, the soft and collinear singularities for real photon
emission are isolated both in the dipole subtraction method following
\citere{Dittmaier:2000mb} and in the phase-space slicing method. For
photons in the initial state the subtraction and slicing variants
described in \citere{Diener:2005me} are applied. The results presented
in the following are obtained with the subtraction method, which
numerically performs better.

The phase-space integration is performed with Monte Carlo techniques.
One of our two codes employs a multi-channel Monte Carlo generator
\cite{Berends:1994pv} similar to the one implemented in {\sc RacoonWW}
\cite{Denner:1999gp,Denner:2002cg}. Our second code uses a different
implementation of a multi-channel Monte Carlo generator with adaptive
weight optimization.

\subsection{Classification of QCD corrections}
\label{se:QCDcorr}

As QCD corrections to Higgs production via VBF we consider the
interference of VBF diagrams of the type shown in \reffi{fig:LOtops}
with the virtual QCD corrections arising from gluon exchange, gluon
fusion, and gluon splitting. We also take into account the
contributions from real gluon emission and gluon-induced processes.
We classify these corrections in the same way as done for the QCD
corrections to $\PH\to4f$ described in \citere{Bredenstein:2006ha}
upon considering possible contributions to the squared amplitude. The
amplitude itself receives contributions from one of the two generic
tree diagrams shown in \reffi{fi:H4f-born-diags} or from both.  Thus,
the square of this amplitude receives contributions from squared and
interference diagrams of the types depicted in
\reffi{fi:H4f-born-ints}.  Type (A) corresponds to the squares of each
of the Born diagrams, type (B) to their interference if two Born
diagrams exist.

After this preliminary consideration we define four different
categories of QCD corrections. Examples of interference diagrams
belonging to these categories are shown in \reffi{fi:virtual-ints},
the corresponding virtual QCD correction diagrams are depicted in
\reffi{fi:virtual}.
\begin{enumerate}
\renewcommand{\labelenumi}{(\alph{enumi})}
\item {\it ``Diagonal'' QCD corrections to squared tree diagrams}
  comprise all interference diagrams resulting from diagram (A) of
  \reffi{fi:H4f-born-ints} by adding one additional gluon.  Cut
  diagrams in which the gluon does not cross the cut correspond to
  virtual one-loop corrections, the ones where the gluon crosses the
  cut correspond to real gluon radiation. Note that interference
  diagrams in which the gluon connects the two closed quark lines
  identically vanish, because their colour structure is proportional
  to $\mathrm{Tr}(\la^h)\mathrm{Tr}(\la^h)=0$, where $\la^h$ is a
  Gell-Mann matrix.  Thus, the only relevant one-loop diagrams in this
  category are gluonic corrections to the $Vq\bar q'$ vertex, as
  illustrated in the first diagram of \reffi{fi:virtual}; the real
  corrections are induced by the corresponding gluon bremsstrahlung
  diagrams.
\bfi
\begin{center}
{\setlength{\unitlength}{.9pt}\SetScale{.9}
\begin{picture}(210,100)
\DashLine( 0,50)(30,50){3}
\Photon(30,50)(60,20){-2}{5}
\Photon(30,50)(60,80){2}{5}
\Vertex(30,50){2.0}
\Vertex(60,80){2.0}
\Vertex(60,20){2.0}
\ArrowLine(60,80)(90,95)
\ArrowLine(90,95)(120,95)
\ArrowLine(120,65)(90,65)
\ArrowLine(90,65)(60,80)
%\SetColor{Red}
\Vertex(90,95){2.0}
\Vertex(90,65){2.0}
\Gluon(90,65)(90,95){2}{4}
\SetColor{Black}
\ArrowLine(120, 5)(60,20)
\ArrowLine(60,20)(120,35)
\DashLine(120,0)(120,100){7}
\ArrowLine(120,95)(150,80)
\ArrowLine(150,80)(120,65)
\ArrowLine(150,20)(120, 5)
\ArrowLine(120,35)(150,20)
\Vertex(180,50){2.0}
\Vertex(150,80){2.0}
\Vertex(150,20){2.0}
\Photon(180,50)(150,20){2}{5}
\Photon(180,50)(150,80){-2}{5}
\DashLine(210,50)(180,50){3}
\put(0,90){{(a)}}
\end{picture}
\hspace*{2em}
\begin{picture}(210,100)
\DashLine(210,50)(180,50){3}
\Photon(180,50)(160,35){-2}{4}
\Photon(180,50)(160,80){2}{5}
\Vertex(180,50){2.0}
\Vertex(160,80){2.0}
\Vertex(160,35){2.0}
\ArrowLine(120,95)(160,80)
\ArrowLine(160,35)(120,65)
\ArrowLine(160,80)(120, 5)
\ArrowLine(120,35)(160,35)
\DashLine(120,0)(120,100){7}
\ArrowLine(60,80)(120,95)
\ArrowLine(90,65)(60,80)
\ArrowLine(120,65)(90,65)
\ArrowLine(120, 5)(60,20)
\ArrowLine(60,20)(90,35)
\ArrowLine(90,35)(120,35)
%\SetColor{Red}
\Vertex(90,35){2.0}
\Vertex(90,65){2.0}
\Gluon(90,35)(90,65){2}{4}
\SetColor{Black}
\Vertex(30,50){2.0}
\Vertex(60,80){2.0}
\Vertex(60,20){2.0}
\Photon(30,50)(60,20){2}{5}
\Photon(30,50)(60,80){-2}{5}
\DashLine(0,50)(30,50){3}
\put(0,90){{(b)}}
\end{picture}
\\[2em]
\begin{picture}(200,100)(15,0)
\ArrowLine(120,95)(150,80)
\ArrowLine(150,80)(120,65)
\ArrowLine(150,20)(120, 5)
\ArrowLine(120,35)(150,20)
\Vertex(180,50){2.0}
\Vertex(150,80){2.0}
\Vertex(150,20){2.0}
\Photon(180,50)(150,20){2}{5}
\Photon(180,50)(150,80){-2}{5}
\DashLine(210,50)(180,50){3}
\DashLine(120,0)(120,100){7}
\DashLine(15,50)(50,50){3}
\Photon(50,50)(70,35){-2}{3}
\Photon(50,50)(70,65){2}{3}
\Vertex(50,50){2.0}
\Vertex(70,35){2.0}
\Vertex(70,65){2.0}
\ArrowLine(70,65)(120, 95)
\ArrowLine(70,35)(85,50)
\ArrowLine(85,50)(70,65)
\ArrowLine(120, 5)(70,35)
\ArrowLine(120,65)(105,50)
\ArrowLine(105,50)(120,35)
%\SetColor{Red}
\Vertex(85,50){2.0}
\Gluon(85,50)(105,50){2}{3}
\Vertex(105,50){2.0}
\SetColor{Black}
\put(8,90){{(c)}}
\end{picture}
\hspace*{2em}
\begin{picture}(200,100)(8,0)
\DashLine(210,50)(180,50){3}
\Photon(180,50)(160,35){-2}{4}
\Photon(180,50)(160,80){2}{5}
\Vertex(180,50){2.0}
\Vertex(160,80){2.0}
\Vertex(160,35){2.0}
\ArrowLine(120,95)(160,80)
\ArrowLine(160,35)(120,65)
\ArrowLine(160,80)(120, 5)
\ArrowLine(120,35)(160,35)
\DashLine(120,0)(120,100){7}
\DashLine(15,50)(50,50){3}
\ArrowLine(70,35)(50,50)
\ArrowLine(50,50)(70,65)
\ArrowLine(70,65)(70,35)
\Vertex(50,50){2.0}
\ArrowLine(95,80)(120,95)
\ArrowLine(120,65)(95,80)
\ArrowLine(95,20)(120,35)
\ArrowLine(120, 5)(95,20)
%\SetColor{Red}
\Vertex(70,35){2.0}
\Vertex(70,65){2.0}
\Gluon(70,65)(95,80){2}{4}
\Gluon(70,35)(95,20){-2}{4}
\Vertex(95,80){2.0}
\Vertex(95,20){2.0}
\SetColor{Black}
\put(8,90){{(d)}}
\end{picture}
}
\vspace*{-1em}
\end{center}
\caption{Categories of interference diagrams contributing to the QCD
  corrections.} 
\label{fi:virtual-ints}
\efi

\bfi
\begin{center}
  \setlength{\unitlength}{1pt}
\begin{picture}(190,100)(-20,10)
\DashLine(15,50)(60,50){3}
\Photon(60,50)(90,20){-2}{5}
\Photon(60,50)(90,80){2}{5}
\Vertex(60,50){2.0}
\Vertex(90,80){2.0}
\Vertex(90,20){2.0}
\ArrowLine(120, 5)( 90,20)
\ArrowLine( 90,20)(120,35)
\ArrowLine( 90,80)(110,90)
\ArrowLine(110,90)(120,95)
\ArrowLine(120,65)(110,70)
\ArrowLine(110,70)(90,80)
\Vertex(110,90){2.0}
\Vertex(110,70){2.0}
\Gluon(110,90)(110,70){2}{3}
\put(-6,47){$\PH$}
\put(62,70){$V$}
\put(62,18){$V$}
\put(125,90){$q_a$}
\put(125,65){$\bar q_b$}
\put(125,30){$q_c$}
\put(125,5){$\bar q_d$}
\put(117,77.5){$\Pg$}
\put(-22,100){(a,b)}
\end{picture}
\begin{picture}(190,100)(-20,10)
\DashLine(15,50)(60,50){3}
\Photon(60,50)(90,20){-2}{5}
\Photon(60,50)(90,80){2}{5}
\Vertex(60,50){2.0}
\Vertex(90,80){2.0}
\Vertex(90,20){2.0}
\ArrowLine(90,80)(120, 95)
\ArrowLine(120,65)(105,72.5)
\ArrowLine(105,72.5)(90,80)
\ArrowLine( 90,20)(105,27.5)
\ArrowLine(105,27.5)(120,35)
\ArrowLine(120, 5)( 90,20)
\Vertex(105,72.5){2.0}
\Vertex(105,27.5){2.0}
\Gluon(105,72.5)(105,27.5){2}{8}
\put(-6,47){$\PH$}
\put(62,70){$V$}
\put(62,18){$V$}
\put(125,90){$q_a$}
\put(125,65){$\bar q_b$}
\put(125,30){$q_c$}
\put(125,5){$\bar q_d$}
\put(112,47.5){$\Pg$}
\put(-22,100){(b)}
\end{picture}\\[2em]
\begin{picture}(190,100)(-20,10)
\DashLine(15,50)(50,50){3}
\Photon(50,50)(70,35){-2}{3}
\Photon(50,50)(70,65){2}{3}
\Vertex(50,50){2.0}
\Vertex(70,35){2.0}
\Vertex(70,65){2.0}
\Vertex(85,50){2.0}
\ArrowLine(70,65)(120, 95)
\ArrowLine(70,35)(85,50)
\ArrowLine(85,50)(70,65)
\ArrowLine(120, 5)(70,35)
\Gluon(85,50)(105,50){2}{3}
\Vertex(105,50){2.0}
\ArrowLine(105,50)(120,65)
\ArrowLine(120,35)(105,50)
\put(-6,47){$\PH$}
\put(52,66){$V$}
\put(52,22){$V$}
\put(125,90){$q_a$}
\put(125,65){$q_c$}
\put(125,30){$\bar q_d$}
\put(125,5){$\bar q_b$}
\put(92,38){$\Pg$}
\put(-22,100){(c)}
\end{picture}
\begin{picture}(190,100)(-20,10)
\DashLine(15,50)(50,50){3}
\ArrowLine(70,35)(50,50)
\ArrowLine(50,50)(70,65)
\ArrowLine(70,65)(70,35)
\Vertex(50,50){2.0}
\Vertex(70,35){2.0}
\Vertex(70,65){2.0}
\Gluon(70,65)(95,80){2}{4}
\Gluon(70,35)(95,20){-2}{4}
\Vertex(95,80){2.0}
\Vertex(95,20){2.0}
\ArrowLine(95,80)(120,95)
\ArrowLine(120,65)(95,80)
\ArrowLine(95,20)(120,35)
\ArrowLine(120, 5)(95,20)
\put(-6,47){$\PH$}
\put(54,66){$Q$}
\put(54,27){$Q$}
\put(76,47){$Q$}
\put(125,90){$q_a$}
\put(125,65){$\bar q_b$}
\put(125,30){$q_c$}
\put(125,5){$\bar q_d$}
\put(78,15){$\Pg$}
\put(78,82){$\Pg$}
\put(-22,100){(d)}
\end{picture}
\end{center}
\caption{Basic diagrams contributing to the virtual QCD corrections to
  $qq\to qq\PH$ where $V=\PW,\PZ$ and $Q=\Pd,\Pu,\Ps,\Pc,\Pb,\Pt$.  The
  categories of QCD corrections, (a)--(d), to which the diagrams
  contribute are indicated.}
\label{fi:virtual}
\efi

Previous calculations
\cite{Spira:1997dg,Han:1992hr,Figy:2003nv,Figy:2004pt,Berger:2004pc}
of NLO QCD corrections focused on this category of corrections to $t$-
and $u$-channel contributions only. This approximation is motivated by
the smallness of $s$-channel contributions, at least in the kinematic
domain relevant for Higgs production via VBF, and by the suppression
of all types of interferences in lowest order. Both of these
suppressions are due to strong enhancements in the $t$- and
$u$-channel weak-boson propagators that receive a small momentum
transfer; only in contributions to squared amplitudes that are related
to squared $t$- and $u$-channel LO graphs four enhancement factors of
this kind can accumulate.  For instance, interferences between two
different $t$- and $u$-channel tree diagrams involve four enhanced
propagators, but they pairwise peak in different regions of phase
space (forward or backward scattered quarks).
\item {\it QCD corrections to interferences} comprise all interference
  diagrams resulting from diagram (B) of \reffi{fi:H4f-born-ints} by
  adding one additional gluon, analogously to the previous category.
  Relevant one-loop diagrams are, thus, vertex corrections or pentagon
  diagrams, as illustrated in the first two diagrams of
  \reffi{fi:virtual}.
\item {\it Corrections induced by one $q\bar q\Pg$ splitting} result
  from loop diagrams exemplified by the third graph in
  \reffi{fi:virtual}. The remaining graphs are obtained by shifting
  the gluon to different positions at the same quark line and by
  interchanging the role of the two quark lines.  Thus, the diagrams
  comprise not only box diagrams but also vertex diagrams.  They do
  not interfere with Born diagrams with the same fermion-number flow
  because of the colour structure, i.e.\ in $\Ord(\als)$ they only
  contribute if two Born diagrams exist.
  
  Some of the squared diagrams of this category actually correspond to
  (collinear-singular) real NLO QCD corrections to loop-induced
  $\PH{+}1$jet production, e.g.\ $qg\to q\PH$ or $q\bar q\to g\PH$.
  Here we consider only the interference contributions of the loop
  diagrams of this category with the lowest-order diagrams where the
  Higgs boson couples to a weak boson (see \reffi{fig:LOtops}),
  resulting in a UV- and IR- (soft and collinear) finite correction.
\item {\it Corrections induced by two $q\bar q\Pg$ splittings} (gg
  fusion) result from diagrams exemplified by the fourth graph in
  \reffi{fi:virtual}.  There are precisely two graphs with opposite
  fermion-number flow in the loop. Again, owing to the colour
  structure (see also below), these diagrams do not interfere with
  Born diagrams with the same fermion-number flow, i.e.\ the existence
  of two Born diagrams is needed.
  
  The squared diagrams of this category actually correspond to
  (collinear-singular) real NNLO QCD corrections to loop-induced Higgs
  production via gluon fusion, $gg\to\PH$. The considered interference
  contributions of the loop diagrams of this category with the
  lowest-order diagrams of \reffi{fig:LOtops}, however, again yield a
  UV- and IR- (soft and collinear) finite correction.
  
  This category of QCD corrections was recently considered in the
  approximation of an infinitely heavy top quark in
  \citere{Andersen:2006ag} and found to be suppressed.  There it was
  also argued that QCD corrections to these small contributions might
  be sizeable, because further gluon exchange between the two incoming
  (anti-)quarks enables an interference with the tree diagram with the
  same fermion-number flow, thereby receiving an enhancement by four
  propagators with small momentum transfer. This contribution has very
  recently been studied in \citere{Andersen:2007mp} and found to be
  completely negligible owing to the appearance of several other
  suppression mechanisms.

\end{enumerate}

\subsection{Structure of virtual corrections}

Since the colour flow in EW loop diagrams is the same as in the
corresponding lowest-order diagrams, the EW one-loop amplitudes
$\A_\EW$ can be decomposed into colour- and CKM-stripped amplitudes
$\M^{VV}_\EW$ exactly in the same way as done in lowest order, where
we decomposed $\A_0$ in terms of $\M_0^{VV}$ \refeq{eq:Mqbarqqbarq}.

According to the above classification, the QCD one-loop amplitudes of
category (a) as well as the vertex corrections of category (b) involve
only the colour operator $C_1$ of \refeq{eq:Ccolour}, while the
pentagon diagrams of category (b) and all loops of categories (c) and
(d) involve only the colour operator $C_2$. Thus, we can decompose the
amplitudes $\A_{\QCD,(a)}$, etc., into colour- and CKM-stripped parts
$\M^{VV}_{\QCD,(a)}$, etc., as follows
\beqar
\A_{\QCD(a)+\QCD(b,\mathrm{vert})}^{\bar u_i d_j\to \bar u_k d_l}
&=& V_{ij} V_{kl}^* C_1^{c_1c_2c'_2c'_1} \M^{\PW\PW}_{\QCD(a)+\QCD(b,\mathrm{vert})}(122'1')
\nn\\ && {}
- \de_{ik}\de_{jl}  C_1^{c_1c'_1c'_2c_2} \M^{\PZ\PZ}_{\QCD(a)+\QCD(b,\mathrm{vert})}(11'2'2),
\nn\\
\A_{\QCD(b,\mathrm{pent})+\QCD(c)+\QCD(d)}^{\bar u_i d_j\to \bar u_k d_l}
&=& V_{ij} V_{kl}^* C_2^{c_1c_2c'_2c'_1} 
\M^{\PW\PW}_{\QCD(b,\mathrm{pent})}(122'1')
\nn\\ && {}
- \de_{ik}\de_{jl}  C_2^{c_1c'_1c'_2c_2} 
\M^{\PZ\PZ}_{\QCD(b,\mathrm{pent})+\QCD(c)+\QCD(d)}(11'2'2),
\nn\\
&\vdots&
\nn\\
\A_{\QCD(a)+\QCD(b,\mathrm{vert})}^{\bar d_i d_j\to \bar d_k d_l}
&=& \de_{ij}\de_{kl} C_1^{c_1c_2c'_2c'_1} 
\M^{\PZ\PZ}_{\QCD(a)+\QCD(b,\mathrm{vert})}(122'1')
\nn\\ && {}
- \de_{ik}\de_{jl}   C_1^{c_1c'_1c'_2c_2} 
\M^{\PZ\PZ}_{\QCD(a)+\QCD(b,\mathrm{vert})}(11'2'2),
\nn\\
\A_{\QCD(b,\mathrm{pent})+\QCD(c)+\QCD(d)}^{\bar d_i d_j\to \bar d_k d_l}
&=& \de_{ij}\de_{kl} C_2^{c_1c_2c'_2c'_1} 
\M^{\PZ\PZ}_{\QCD(b,\mathrm{pent})+\QCD(c)+\QCD(d)}(122'1')
\nn\\ && {}
- \de_{ik}\de_{jl}   C_2^{c_1c'_1c'_2c_2} 
\M^{\PZ\PZ}_{\QCD(b,\mathrm{pent})+\QCD(c)+\QCD(d)}(11'2'2).
\hspace{2em}
\eeqar
Note that W-mediated parts $\M^{\PW\PW}$ do not receive contributions
of categories (c) and (d).

Since the lowest-order amplitudes only involve colour operators $C_1$,
the following colour sums appear in the calculation of squared
lowest-order amplitudes and of interferences between one-loop and
lowest-order matrix elements:
\beq
\begin{array}[b]{rclcrcl}
X^{(A)}_1 &=& \disp\sum_{a,b,c,d}(C^{abcd\,*}_1C^{abcd}_1) =
(N^{\mathrm{c}})^2,
&\qquad& X^{(A)}_2 &=& \disp\sum_{a,b,c,d}(C^{abcd\,*}_1C^{abcd}_2) =0,
\\[1.3em]
X^{(B)}_1 &=& \disp\sum_{a,b,c,d}(C^{abcd\,*}_1C^{adcb}_1) =
N^{\mathrm{c}},
&& X^{(B)}_2 &=& \disp\sum_{a,b,c,d}(C^{abcd\,*}_1C^{adcb}_2) =
N^{\mathrm{c}},
\end{array}
\eeq
where $\sum_{a,b,c,d}$ stands for the sum over the colour indices
$a,b,c,d$, and $N^{\mathrm{c}}=3$ is the colour factor for a quark.
Squared Born diagrams, as illustrated in type (A) of
\reffi{fi:H4f-born-diags}, are proportional to $X^{(A)}_1$,
lowest-order interference diagrams of type (B) are proportional to
$X^{(B)}_1$.  The situation is analogous for all EW one-loop diagrams.
By definition, category (a) of the gluonic diagrams comprises all
one-loop QCD corrections proportional to $X^{(A)}_1$.  In category
(b), the vertex corrections are proportional to $X^{(B)}_1$ and the
pentagons to $X^{(B)}_2$.  Categories (c) and (d) receive only
contributions from $X^{(B)}_2$; interferences of one-loop diagrams
like (c) and (d) in \reffi{fi:virtual} with Born diagrams of the same
fermion-number flow vanish because of $X^{(A)}_2=0$.  Finally, for the
one-loop corrections to the squared matrix elements we obtain
\beqar
\lefteqn{
\sum_{\{c_i,c'_i\}} 2\Re\left\{
(\A_{0}^{\bar u_i d_j\to \bar u_k d_l})^*
\A_{1}^{\bar u_i d_j\to \bar u_k d_l} \right\}
}
\hspace*{1em}
&& 
\nn\\
&=&
2\Re\Bigl\{
V_{ij}^* V_{kl} \M^{\PW\PW}_0(122'1')^* \Bigl[
9 V_{ij} V_{kl}^* \M^{\PW\PW}_{\EW+\QCD(a)}(122'1')
\nn\\
&& \hspace*{5em}
-3 \de_{ik}\de_{jl}\M^{\PZ\PZ}_{\EW+\QCD(b)+\QCD(c)+\QCD(d)}(11'2'2)
\Bigr]
\nn\\
&& \phantom{2\Re\Bigl\{}
+\de_{ik}\de_{jl}\M^{\PZ\PZ}_0(11'2'2)^* \Bigl[
9\M^{\PZ\PZ}_{\EW+\QCD(a)}(11'2'2)
-3 V_{ij} V_{kl}^* \M^{\PW\PW}_{\EW+\QCD(b)}(122'1')
\Bigr]
\Bigr\},
\nn\\
&\vdots&
\nn\\
\lefteqn{
\sum_{\{c_i,c'_i\}} 2\Re\left\{
(\A_{0}^{\bar d_i d_j\to \bar d_k d_l})^*
\A_{1}^{\bar d_i d_j\to \bar d_k d_l} \right\}
}
\hspace*{1em}
&& 
\nn\\
&=&
2\Re\Bigl\{
\de_{ij}\de_{kl}\M^{\PZ\PZ}_0(122'1')^* \Bigl[
9\M^{\PZ\PZ}_{\EW+\QCD(a)}(122'1')
\nn\\
&& \hspace*{5em}
- 3\de_{ik}\de_{jl}  
\M^{\PZ\PZ}_{\EW+\QCD(b)+\QCD(c)+\QCD(d)}(11'2'2) \Bigr]
\nn\\
&& \phantom{2\Re\Bigl\{}
+\de_{ik}\de_{jl} \M^{\PZ\PZ}_0(11'2'2)^* \Bigl[
9\M^{\PZ\PZ}_{\EW+\QCD(a)}(11'2'2)
\nn\\
&& \hspace*{5em}
- 3\de_{ij}\de_{kl}
\M^{\PZ\PZ}_{\EW+\QCD(b)+\QCD(c)+\QCD(d)}(122'1') \Bigr]
\Bigr\}.
\eeqar
As already observed for the squared LO amplitudes, also here only
absolute values of the CKM matrix elements, such as $|V_{ij}|^2$,
contribute after contracting the Kronecker deltas of the generation
indices.

\subsection{Hadronic cross section}
\label{se:hadxsection}

The hadronic cross section $\sigma_{\Pp\Pp}(P_1,P_2)$ for colliding
protons results from the partonic cross section
$\hat\sigma_{ij}(x_1P_1,x_2P_2)$ upon convolution with the parton
densities $f_i(x_l,\muF^2)$, which corresponds to parton $i$ carrying
the fraction $x_l$ of the proton momentum $P_l$ ($l=1,2$),
\beq
\sigma_{\Pp\Pp}(P_1,P_2) = 
\int_0^1\rd x_1\, \int_0^1\rd x_2\, \sum_{i,j} f_i(x_1,\muF^2) f_j(x_2,\muF^2)
\hat\sigma_{ij}(x_1P_1,x_2P_2),
\eeq
where $\muF$ is the factorization scale that separates the hard
partonic process from the soft physics contained in the PDFs. The sum
over the partons $i,j$ includes all quarks, antiquarks, gluons, and
the photon. In LO only quarks and/or antiquarks are present in the
initial state, in NLO also processes with one gluon or photon
contribute.  In detail the NLO parton cross sections read
\beqar
\hat\sigma_{qq} &=& \hat\sigma_{qq,\LO} + \hat\sigma_{qq,\virt} + 
\hat\sigma_{qq,\real} + \hat\sigma_{qq,\fact},
\nn\\
\hat\sigma_{q\Pg} &=&
\hat\sigma_{q\Pg,\real} + \hat\sigma_{q\Pg,\fact}, \qquad
\hat\sigma_{\Pg q} =
\hat\sigma_{\Pg q,\real} + \hat\sigma_{\Pg q,\fact},
\nn\\
\hat\sigma_{q\ga} &=&
\hat\sigma_{q\ga,\real} + \hat\sigma_{q\ga,\fact},\qquad
\hat\sigma_{\ga q} =
\hat\sigma_{\ga q,\real} + \hat\sigma_{\ga q,\fact},
\nn\\
\hat\sigma_{\Pg\Pg} &=& \hat\sigma_{\Pg\ga} = \hat\sigma_{\ga\Pg} = 
\hat\sigma_{\ga\ga} = 0,
\eeqar
where $q$ generically stands for any relevant quark or antiquark.  The
LO and virtual one-loop contributions (``LO'' and ``virt'') involve
the partonic $2\to2$ kinematics, while real emission contributions
(``real'') are of the type $2\to3$ with one additional light
(anti)quark, gluon, or photon in the final state.  The calculation of
these subcontributions has been briefly described in the previous
sections. The contribution called ``fact'' results from the PDF
redefinition necessary to absorb collinear initial-state singularities
into the PDFs via factorization, so that the partonic cross sections
$\hat\sigma_{ij}$ are free of such singularities. This separation
introduces a logarithmic dependence on the factorization scale $\muF$
in $\hat\sigma_{ij,\fact}$ that compensates the implicit $\muF$
dependence in the PDFs in NLO accuracy.  The factorization explicitly
proceeds as follows.

\begin{sloppypar}
  The virtual and real contributions of the parton cross sections
  contain mass singularities of the form $\alpha_{\mathrm{s}}\ln(m_q)$
  and $\alpha\ln(m_q)$, which are due to collinear gluon/photon
  radiation off the initial-state quarks or due to a collinear
  splitting $\Pg/\gamma\to q\bar q$ of initial-state gluons or
  photons.  For processes that in LO involve only quarks and/or
  antiquarks in the initial state, the factorization is achieved by
  replacing the (anti-)quark distribution $f_q(x)$ according to (see
  e.g. \citere{Diener:2005me})
\beqar
\lefteqn{f_q(x,\muF^2) \to f_q(x,\muF^2) 
-\int_x^1 \frac{\rd z}{z} \, f_q\biggl(\frac{x}{z},\muF^2\biggr)} &&
\nn\\
&& \quad {} 
\times
\biggl\{
\frac{2\alpha_{\mathrm{s}}}{3\pi} 
\biggl(
\ln\biggl(\frac{\muF^2}{m_q^2}\biggr) \Bigl[ P_{ff}(z) \Bigr]_+
-\Bigl[ P_{ff}(z) \Bigl(2\ln(1-z)+1\Bigr) \Bigr]_+
+C_{ff,\QCD}(z)
\biggr)
\biggr\}
\nn\\
&& \quad {} 
\phantom{\times \biggl\{}
+\frac{\al}{2\pi} \, Q_q^2 
\biggl(
\ln\biggl(\frac{\muF^2}{m_q^2}\biggr) \Bigl[ P_{ff}(z) \Bigr]_+
-\Bigl[ P_{ff}(z) \Bigl(2\ln(1-z)+1\Bigr) \Bigr]_+
+C_{ff,\QED}(z)
\biggr)
\biggr\}
\nn\\
&& {}
-\int_x^1 \frac{\rd z}{z} \, f_\Pg\biggl(\frac{x}{z},\muF^2\biggr)
\frac{\als}{4\pi} 
\biggl\{
\ln\biggl(\frac{\muF^2}{m_q^2}\biggr) P_{f\gamma}(z)
+C_{f\Pg}(z)
\biggr\}
\nn\\
&& {}
-\int_x^1 \frac{\rd z}{z} \, f_\gamma\biggl(\frac{x}{z},\muF^2\biggr)
\frac{\alpha}{2\pi} \, 3Q_q^2 
\biggl\{
\ln\biggl(\frac{\muF^2}{m_q^2}\biggr) P_{f\gamma}(z)
+C_{f\gamma}(z)
\biggr\},
\label{eq:factorization}
\eeqar
where $C_{ij}(z)$ are the so-called coefficient functions, and the
splitting functions $P_{ij}(z)$ are defined as
\beq
P_{ff}(z) = \frac{1+z^2}{1-z}, \qquad
P_{f\gamma}(z) = z^2+(1-z)^2.
\label{eq:splittings}
\eeq
Starting from the hadronic LO cross section after the substitution
\refeq{eq:factorization}, the factorization contributions
$\hat\sigma_{ij,\fact}$ correspond to the terms of ${\cal
  O}(\alpha_{\mathrm{s}})$ and ${\cal O}(\alpha)$ involving the PDF of
$i$ and $j$.  The replacement \refeq{eq:factorization} defines the
same finite coefficient functions as the usual $D$-dimensional
regularization for exactly massless partons where the $\ln(m_q)$ terms
appear as $1/(D-4)$ poles.  The actual form of the coefficient
functions defines the finite parts of the NLO corrections and, thus,
the factorization scheme.  Following standard definitions of QCD, we
distinguish the $\MSbar$ and DIS-like schemes which are formally
defined by
\beqar
C^{\MSbar}_{ff}(z) &=& 
C^{\MSbar}_{f\Pg}(z) = C^{\MSbar}_{f\gamma}(z) = 0,
\nn\\
C^{\DIS}_{ff}(z) &=& 
\left[ P_{ff}(z)\left(\ln\biggl(\frac{1-z}{z}\biggr)-\frac{3}{4}\right)
+\frac{9+5z}{4} \right]_+,
\nn\\
C^{\DIS}_{f\Pg}(z) &=& C^{\DIS}_{f\gamma}(z) = 
P_{f\gamma}(z)\ln\biggl(\frac{1-z}{z}\biggr) -8z^2+8z-1.
\label{eq:PDFcoeff}
\eeqar
The $\MSbar$ scheme is motivated by formal simplicity, because it
merely rearranges the IR-divergent terms (plus some trivial constants)
as defined in dimensional regularization.  The DIS-like scheme is
defined in such a way that the deep inelastic scattering (DIS)
structure function $F_2$ does not receive any corrections; in other
words, the radiative corrections to electron--proton DIS are
implicitly contained in the PDFs.
\end{sloppypar}

Whatever scheme has been adopted in the extraction of PDFs from
experimental data, the same scheme has to be used when predictions for
other experiments are made using these PDFs.  In particular, the
absorption of the collinear singularities of both QCD and QED origin
into PDFs requires the inclusion of the corresponding QCD and QED
corrections into the
Dok\-shitzer--Gri\-bov--Lipa\-tov--Alta\-relli--Pa\-risi (DGLAP)
evolution of these distributions and into their fit to experimental
data.  We use the MRST2004QED PDFs \cite{Martin:2004dh} which
consistently include QCD and QED NLO corrections. These PDFs include a
photon distribution function for the proton and thus allow to take
into account photon-induced partonic processes. As explained in
\citere{Diener:2005me}, the consistent use of these PDFs requires the
$\MSbar$ factorization scheme for the QCD corrections, but the DIS
scheme for the QED corrections, i.e.\ we employ $C^{\MSbar}_{ff}$ and
$C^{\MSbar}_{f\Pg}$ of \refeq{eq:PDFcoeff} for the QCD, but
$C^{\DIS}_{ff}$ and $C^{\DIS}_{f\gamma}$ for the QED corrections.

\section{Numerical results}
\label{se:numres}

\subsection{Input parameters and setup}
\label{se:input}

We use the following set of input parameters \cite{Eidelman:2004wy},
\beqar
\begin{array}[b]{r@{\,}lr@{\,}lr@{\,}l}
G_{\mu} &= 1.16637\times 10^{-5}\GeV^{-2}, \quad &
\al(0) &= 1/137.03599911, \quad 
& \alpha_{\mathrm{s}}(\MZ) &= 0.1187,\\                          %2004
\MW^{\LEP} &= 80.425\GeV, & \GW^{\LEP} &= 2.124\GeV, && \\  % 2004
\MZ^{\LEP} &= 91.1876\GeV,& \GZ^{\LEP} &= 2.4952\GeV, && \\
\Me &= 0.51099892 \MeV, & \Mmy &= 105.658369 \MeV, 
&\Mta &= 1.77699\GeV,\\
\Mu &= 66 \MeV, & \Mc &=1.2 \GeV, 
& \Mt &= 174.3\GeV, \\                                      % no PDG
\Md &= 66 \MeV, & \Ms &=150 \MeV, & \Mb &= 4.3\GeV.
\end{array}
\eeqar
If not stated otherwise, the Higgs-boson mass is set to 
\beq
\MH=120\GeV.
\eeq
Using the complex-mass scheme \cite{Denner:2005es}, we employ a fixed
width in the resonant W- and Z-boson propagators in contrast to the
approach used at LEP to fit the W~and Z~resonances, where running
widths are taken.  Therefore, we have to convert the ``on-shell''
values of $M_V^{\LEP}$ and $\Ga_V^{\LEP}$ ($V=\PW,\PZ$), resulting
from LEP, to the ``pole values'' denoted by $M_V$ and $\Ga_V$. The
relation between the two sets of values is given by
\cite{Bardin:1988xt}
\beq\label{eq:m_ga_pole}
M_V = M_V^{\LEP}/
\sqrt{1+(\Ga_V^{\LEP}/M_V^{\LEP})^2},
\qquad
\Ga_V = \Ga_V^{\LEP}/
\sqrt{1+(\Ga_V^{\LEP}/M_V^{\LEP})^2},
\eeq
leading to
\beqar
\begin{array}[b]{r@{\,}l@{\qquad}r@{\,}l}
\MW &= 80.397\ldots\GeV, & \GW &= 2.123\ldots\GeV, \\
\MZ &= 91.1535\ldots\GeV,& \GZ &= 2.4943\ldots\GeV.
\label{eq:m_ga_pole_num}
\end{array}
\eeqar
We make use of these mass parameters in the numerics discussed below,
although the difference between using $M_V$ or $M_V^{\LEP}$ would be
hardly visible.  

The masses of the light quarks are adjusted to reproduce the hadronic
contribution to the photonic vacuum polarization of
\citere{Jegerlehner:2001ca}. Since quark mixing effects are
suppressed%
\footnote{We checked that the cross section without cuts changes by
  one per mille and the one with VBF cuts by less than 0.01\% when
  using a realistic quark mixing matrix.}  we neglect quark mixing and
use a unit CKM matrix.

We use the $\GF$ scheme, \ie we derive the electromagnetic coupling
constant from the Fermi constant according to 
\beq
 \alpha_{\GF} = \sqrt{2}\GF\MW^2(1-\MW^2/\MZ^2)/\pi.
\eeq
In this scheme, the weak corrections to muon decay $\De r$ are
included in the charge renormalization constant (see \eg
\citere{Ciccolini:2003jy}).  As a consequence, the EW corrections are
practically independent of the masses of the light quarks. Moreover,
this definition effectively resums the contributions associated with
the running of $\al$ from zero to the W-boson mass and absorbs leading
universal corrections $\propto\GF\Mt^2$ from the $\rho$~parameter into
the LO amplitude.

We use the MRST2004QED PDFs \cite{Martin:2004dh} which consistently
include $\Oa$ QED corrections.  Since no associated LO PDFs exist, we
use these distributions both for LO and NLO predictions.  We do not
include processes with external bottom quarks in our default set-up.
These are suppressed either because of the smallness of the b-quark
densities or due to $s$-channel suppression.  Partonic processes
involving b~quarks are, however, included in our code in LO.  As
discussed in \refse{se:b-contributions}, these contributions are at the
level of a few per cent.  In contrast to \citere{Ciccolini:2007jr}, we
use $\MW$ (instead of $\MH$) as factorization scale both for QCD and
QED collinear contributions, which is a better scale choice when
considering large Higgs-boson masses.  For the calculation of the
strong coupling constant we employ $\MW$ as the default
renormalization scale, include 5 flavours in the two-loop running, and
fix $\alpha_{\mathrm{s}}(\MZ)= 0.1187$.

Jet reconstruction from final-state partons is performed using the
$k_{\rT}$-algorithm \cite{Catani:1992zp} as described in
\citere{Blazey:2000qt}. Jets are reconstructed from partons of
pseudorapidity $|\eta|<5$ using a jet resolution parameter $D=0.8$.
Real photons are recombined with jets according to the same algorithm.
Thus, in real photon radiation events, final states may consist of
jets plus a real identifiable photon, or of jets only.

We study total cross sections and cross sections for the set of
experimental ``VBF cuts'' defined in \citere{Figy:2004pt}. These cuts
are expected to significantly suppress backgrounds to VBF processes,
enhancing the signal-to-background ratio. We require at least two hard
jets with
\beq\label{taggingjets}
p_{\rT\rj}>20\GeV,\qquad |y_{\rj}|<4.5,
\eeq
where $p_{\rT\rj}$ is the transverse momentum of the jet and $y_{\rj}$
its rapidity. Two tagging jets $\rj_1$ and $\rj_2$ are defined as the
two jets passing the cuts \refeq{taggingjets} with highest $p_{\rT}$
such that $p_{\rT\rj_1}>p_{\rT\rj_2}$. Furthermore, we require that
the tagging jets have a large rapidity separation and reside in
opposite detector hemispheres:
\beq
\De y_{\rj\rj}\equiv | y_{\rj_1}-y_{\rj_2}| > 4, 
\qquad y_{\rj_1}\cdot y_{\rj_2}< 0.
\eeq

All presented results have been obtained using the subtraction method.
For the results in the tables we used $10^8$ events for the setup with
VBF cuts and $5\times10^7$ events without cuts. For the plots of $\MH$
and factorization scale dependence we generated $10^7$ events without
cuts and $2\times10^7$ events with VBF cuts. The plots for the
distributions are based on $10^9$ events. Generally, the real
corrections and the finite virtual QCD corrections are only calculated
for each 10th event, the finite virtual EW corrections only for each
100th event.

\subsection{Results for integrated cross sections}

\begin{figure}[t]
%\framebox{
\includegraphics[bb= 85 440 285 660, scale=1]{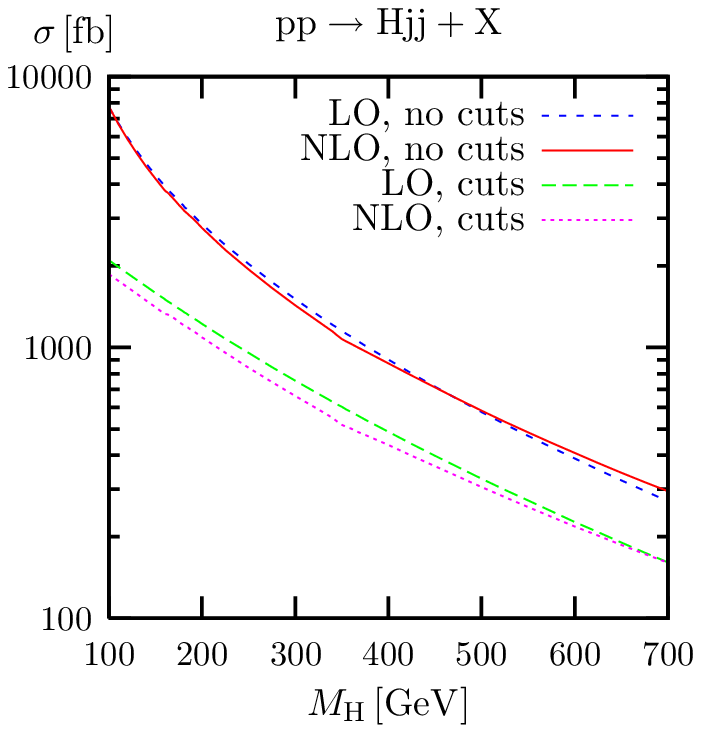}
%}
\quad
%\framebox{
\includegraphics[bb= 85 440 285 660, scale=1]{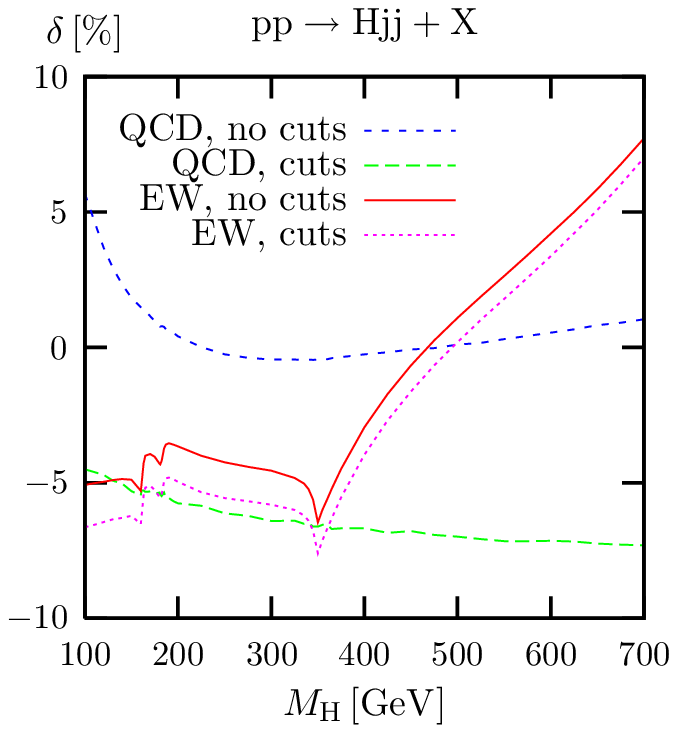}
%}
\caption{Higgs-boson-mass dependence of LO and complete NLO
  cross section (left) and relative EW and QCD corrections (right)
  without and with VBF cuts.}
\label{fi:mhdep_hh}
\end{figure}

We first consider results for integrated cross sections.  In
\reffi{fi:mhdep_hh} we plot the total cross section with and without
VBF cuts as a function of the Higgs-boson mass. In the left panel we
show the absolute predictions in LO and in NLO including QCD and EW
corrections. For $\MH=100\GeV$ to about $200\GeV$ the results without
cuts are larger by a factor 2--4, while for $\MH=700\GeV$ this factor
reduces to 1.7. In the right panel we show the relative QCD and EW
corrections separately. For Higgs-boson masses in the range
$100$--$200\GeV$, without cuts, the QCD corrections drop from +5\% to
0\%, and the EW corrections are about $-5\%$ depending only weakly on
the Higgs-boson mass. For very small Higgs-boson masses, QCD and EW
corrections cancel each other substantially.  With VBF cuts the EW
corrections are somewhat more negative, while the QCD corrections vary
between $-4\%$ and $-6\%$.  For higher Higgs-boson masses, the QCD
corrections do not change much and reach $1\%$ and $-7\%$ at
$\MH=700\GeV$ without cuts and with VBF cuts, respectively.  The EW
corrections increase steadily with the Higgs-boson mass up to $8\%$
and $7\%$ at $\MH=700\GeV$ without cuts and with VBF cuts,
respectively.  In the EW corrections the WW, ZZ, and tt thresholds are
clearly visible.

It is interesting to note that, at least for Higgs-boson masses below
$200\GeV$, the EW corrections to the full VBF channel are similar in
size and sign to the subreactions $\Pp\Pp\to\PW\PH/\PZ\PH+X$
\cite{Ciccolini:2003jy}. Compared to the related decays
$\PH\to\PW\PW/\PZ\PZ\to4f$
\cite{Bredenstein:2006rh,Bredenstein:2006ha} the size is similar, but
for low Higgs masses of $100{-}200\GeV$ the sign is different.

In \refta{ta:xsection_nocuts} we present numbers for integrated cross
sections for $\MH=120$, 150, 200, 400, and $700\GeV$ without any cuts and
in \refta{ta:xsection_cuts} results for the VBF cuts defined above.
\begin{table}
\def\phm{\phantom{-}}
\def\phn{\phantom{0}}
\def\phe{\phantom{(0)}}
\centerline{
\begin{tabular}{|c|c|c|c|c|c|c|}
\hline
$\MH\ [\GeV]$ & 120 & 150 & 200 & 400& 700\\
\hline
$\si_{\mathrm{LO}}\ [\fb]$ 
& 5943(1)
& 4331(1)
& 2855.4(6) % 
&  900.7(1) % 
&  270.51(4) %
\\
$\si_{\mathrm{NLO}}\ [\fb]$ 
& 5872(2)
& 4202(2)
& 2765(1)
&  871.8(3)
&  294.33(9) %
\\
$\de_{\mathrm{EW}}\ [\%]$ 
& $-4.94(2)$
& $-4.91(2)$
& $-3.67(1)$
& $-2.97(1)$
& $\phm7.74(2)$ %
\\
$\de_{\EW,qq}\ [\%]$ 
& $-5.79(2)$
& $-5.92(2)$
& $-4.85(1)$
& $-4.50(1)$
& $\phm5.99(2)$ %
\\
$\de_{\gammainduced}\ [\%]$ 
&  $\phm0.85\phe$ % 0.8484(8)
&  $\phm1.00\phe$ % 1.001(1)
&  $\phm1.18\phe$ % 1.184(1)
&  $\phm1.53\phe$ % 1.530(1)
&  $\phm1.75\phe$ % 1.750(1) 
\\
$\de_{\mathrm{QCD}}\ [\%]$ 
& $\phm3.75(5)$
& $\phm1.94(3)$
& $\phm0.49(3)$
& $-0.24(3)$ 
& $\phm1.06(3)$ %
\\
$\de_{\mathrm{QCD,diag}}\ [\%]$ 
& $\phm3.97(3)$   % 
& $\phm2.04(3)$   % 
& $\phm0.55(3)$   % 
& $-0.06(3)$   % 
& $\phm1.14(3)$   % 
\\
$\de_{\mathrm{QCD,nondiag}}\ [\%]$ 
&$\phm0.010(2)$  % 
&$\phm0.027(2)$  % 
&$\phm0.050(1)$  % 
&$\phm0.026\phe$ % $\phm0.0258(4)$
&$\phm0.013\phe$ % $\phm0.0131(1)$ %
\\
$\de_{\gsplit}\ [\%]$ 
&$-0.015(1)$      % $-0.0153(5)$
&$\phm 0.059(1)$  % $\phm 0.0587(7)$
&$\phm 0.110(1)$ % $\phm 0.1095(9)$
&$\phm 0.040(1)$ % $\phm 0.0402(3)$
&$\phm 0.017(1)$ % $\phm 0.0174(2)$ 
\\
$\de_{\ggfus}\ [\%]$ 
& $-0.19(1)$ % $-0.194(2)$
& $-0.20\phn$ % $-0.200(2)$
& $-0.22\phn$ % $-0.219(2)$
& $-0.24\phn$ % $-0.242(1)$
& $-0.11(1)$ % $-0.1145(8)$ %
\\
$\de_{\hhtwoloops}\ [\%]$ 
& $\phm0.0027\phn$ % $0.002734(8)$
& $\phm0.0073\phn$  % $0.00728(2)$
& $\phm0.025\phn$  % $0.02468(6)$
& $\phm0.42\phn$   % $0.4227(9)$
& $\phm4.03(1)$    % $4.026(8)$
\\
\hline
\end{tabular}
}
\caption{Cross section for $\ppjjh$ in LO and NLO without cuts
  and relative EW and
  QCD corrections split into various subcontributions.}
\label{ta:xsection_nocuts}
%\end{table}
%
\vspace{2em}
%\begin{table}
\def\phm{\phantom{-}}
\def\phn{\phantom{0}}
\def\phe{\phantom{(0)}}
\centerline{
\begin{tabular}{|c|c|c|c|c|c|}
\hline
$\MH\ [\GeV]$ & 120 & 150 & 200 & 400& 700\\
\hline
$\si_{\mathrm{LO}}\ [\fb]$ 
& 1876.3(5) % 
& 1589.8(4) % 
& 1221.1(3) % 
&  487.31(9)
& 160.67(2) % 
\\
$\si_{\mathrm{NLO}}\ [\fb]$ 
& 1665(1) % 
& 1407.5(8) % 
& 1091.3(5) % 
&  435.4(2)
&  160.36(5) % 
\\
$\de_{\mathrm{EW}}\ [\%]$ 
& $-6.47(2)$  % 
& $-6.27(2)$  % 
& $-4.98(1)$  % 
& $-3.99(1)$  %
& $\phm 6.99(2)$  % 
\\
$\de_{\EW,qq}\ [\%]$ 
& $-7.57(2)$  % 
& $-7.42(2)$  % 
& $-6.19(1)$  % 
& $-5.37(1)$  %
& $\phm 5.44(2)$  % 
\\
$\de_{\gammainduced}\ [\%]$ 
& $\phm1.10\phe$ % 1.102(1)  % 
& $\phm1.15\phe$ % 1.151(1)  % 
& $\phm1.22\phe$ % 1.216(1)  % 
& $\phm1.38\phe$ % 1.384(1)  %
& $\phm1.55\phe$ % 1.547(1)  % 
\\
$\de_{\mathrm{QCD}}\ [\%]$ 
& $-4.77(4)$  % 
& $-5.20(4)$  % 
& $-5.65(3)$ % 
& $-6.67(3)$ 
& $-7.18(2)$
\\
$\de_{\mathrm{QCD,diag}}\ [\%]$ 
& $-4.75(4)$  % 
& $-5.17(4)$  % 
& $-5.66(4)$  % 
& $-6.63(3)$  %
& $-7.18(2)$  %
\\
$\de_{\mathrm{QCD,nondiag}}\ [\%]$ 
&$-0.011\phe$  % $-0.0112(2)$  % 
&$-0.0052(1)$  % $-0.0052(1)$  % 
&$\phm0.0032(1)$ % $\phm0.00323(8)$ % 
&$\phm0.0030\phe$  % $\phm0.00297(2)$
&$\phm0.0022\phe$ % $\phm0.00221(1)$ %
\\
$\de_{\gsplit}\ [\%]$ 
&$   -0.0085(1)$ % $   -0.00850(7)$ % 
&$\phm0.0084(1)$ % $\phm0.00843(9)$ % 
&$\phm0.027\phe$ % $\phm0.0272(2)$  % 
&$\phm0.014\phe$ % $\phm0.01418(5)$
&$0.0074\phn$    % $0.00737(2)$ %
\\
$\de_{\ggfus}\ [\%]$ 
& $-0.030\phe$ %  $-0.0303(2)$ % 
& $-0.030\phe$ %  $-0.0297(2)$ % 
& $-0.028(1)$ %  $-0.0275(1)$ % 
& $-0.020\phe$ % $-0.02004(6)$
& $-0.014\phe$ % $-0.01387(3)$ %
\\
$\de_{\hhtwoloops}\ [\%]$ 
& $\phm0.0035\phe$ % $0.00350(1)$
& $\phm0.0086(1)$ % $0.00855(2)$
& $\phm0.027\phe$ % $0.02703(7)$
& $\phm0.43\phe$  % $0.4328(9)$
& $\phm4.06(1)$   % $4.060(7)$
\\
\hline
\end{tabular}
}
\caption{As in \refta{ta:xsection_nocuts}, but with VBF cuts applied.}
%\caption{Cross section for $\ppjjh$ in LO and NLO
%  and relative EW and
%  QCD corrections split into various subcontributions,  with VBF cuts.}
\label{ta:xsection_cuts}
\end{table}
We list the LO cross section $\si_{\mathrm{LO}}$, the cross section
including NLO QCD and EW corrections, $\si_{\mathrm{NLO}}$, and
various contributions to the relative corrections.  The complete EW
corrections $\de_{\mathrm{EW}}$ comprise the EW corrections resulting
from loop diagrams and real photon radiation,
$\de_{\EW,qq}$, and the corrections from photon-induced processes
$\de_{\gammainduced}$. Furthermore, $\de_{\EW,qq}$ includes the
dominant two-loop correction $\de_{\hhtwoloops}$ due to Higgs-boson
self-interaction, which was introduced in \refse{se:nlocorr}.  The QCD
corrections $\de_{\mathrm{QCD}}$ are decomposed in the diagonal
contributions $\de_{\mathrm{QCD,diag}}$, non-diagonal contributions
$\de_{\mathrm{QCD,nondiag}}$, the contribution resulting from gluon
splitting $\de_{\gsplit}$, and those from gluon-gluon fusion
$\de_{\mathrm{\ggfus}}$, as explained in \refse{se:QCDcorr}.

The QCD corrections are dominated by the diagonal contributions, \ie
by the vector-boson--quark--antiquark vertex corrections to squared LO
diagrams. All other contributions are at the per-mille level and even
partially cancel each other.  They are not enhanced by contributions
of two $t$- or $u$-channel vector bosons with small virtuality and
therefore even further suppressed when applying VBF cuts.  The
photon-induced EW corrections are about $1\%$ and reduce the
EW corrections for small and intermediate $\MH$.  The
two-loop correction $\de_{\hhtwoloops}$ is negligible in the low-$\MH$
region, but becomes important for large Higgs-boson masses. For
$\MH=700\GeV$ this contribution yields $+4\%$ and constitutes about
$50\%$ of the total EW corrections.  Obviously for Higgs masses in
this region and above the perturbative expansion breaks down, and the
two-loop factor $\de_{\hhtwoloops}$ might serve as an estimate of the
theoretical uncertainty.

\subsection{\boldmath{Subcontributions from $s$ channel and 
    $t/u$ interference}}
\label{se:further-split}

Previous calculations of the VBF process
\cite{Spira:1997dg,Han:1992hr,Figy:2003nv,Figy:2004pt,Berger:2004pc}
have consistently neglected $s$-channel contributions (``Higgs
strahlung''), which involve diagrams where one of the vector bosons
can become resonant, as well as the interference between $t$- and
$u$-channel fusion diagrams. To better understand the effect of these
approximations we have calculated these contributions to the
integrated cross section. In \refta{ta:splits_nocuts} and
\refta{ta:splits_cuts} we present, with and without VBF cuts,
respectively, contributions from $s$-channel processes, $\si_s$, and
from $t$/$u$-channel interference terms $\si_{\mbox{\scriptsize
    $t/u$-int}}$, at both LO and NLO.
\begin{table}
\def\phm{\phantom{-}}
\def\phn{\phantom{0}}
\def\phe{\phantom{(0)}}
\centerline{
\begin{tabular}{|c|c|c|c|c|c|}
\hline
$\MH\ [\GeV]$ & 120 & 150 &  200 & 400 & 700\\
\hline
$\si_{\LO,s}\ [\fb]$ 
& 1294.4(2)
&  639.4(1)
&  244.26(4)
&   19.69\phe  %  19.690(3)
&    2.11\phe$\!\!\!$ %   2.1101(3)
\\
$\si_{\NLO,s}\ [\fb]$ 
& 1582.1(4)
&  769.4(2)
&  289.80(9)
&   21.72(1)  %   21.716(8)
&    2.29(1)$\!\!\!$ %   2.2950(7)
\\
$\si_{\LO,\mbox{\scriptsize $t/u$-int}}\ [\fb]$ 
& ${-9.2}\phe$ % $-9.246(2)$
& ${-5.6}\phe$ % $-5.607(2)$
& ${-2.71}\phe$ % $-2.707(1)$
& ${-0.32}\phe$ % $-0.3228(2)$
& ${-0.041}\phe\!\!\!$% $-0.04060(3)$
\\
$\si_{\NLO,\mbox{\scriptsize $t/u$-int}}\ [\fb]$ 
& $-27.6\phe$ % $-27.57(2)$
& $ -9.4\phe$ % $ -9.42(1)$
& $  0.04(1)$ % $  0.04(1)$
& $-1.08(1)$  % $-1.076(2)$
& $-0.19\phe\!\!\!$ % $-0.1892(5)$
\\
\hline
\end{tabular}
}
\caption{$s$-channel contributions and contributions from interference
  between $t$ and $u$ channels to the $\ppjjh$ cross section at LO and
  NLO, without any cuts.} 
\label{ta:splits_nocuts}
%\end{table}
%%
%\begin{table}
\vspace{2em}
\def\phm{\phantom{-}}
\def\phn{\phantom{0}}
\def\phe{\phantom{(0)}}
\centerline{
\begin{tabular}{|c|c|c|c|c|c|}
\hline
$\MH\ [\GeV]$ & 120 & 150 & 200 & 400& 700\\
\hline
$\si_{\LO,s}\ [\fb]$ 
& $0.0025$   %  0.002524(4)
& $0.0015$   % 0.001507(2)
& $0.00071$  % 0.0007101(7)
& $0.000072$ % 0.00007211(4)
& $0.0000069$ % 0.000006917(3)
\\
$\si_{\NLO,s}\ [\fb]$ 
& $\phm 9.45(1)$  
& $\phm5.21(1)$ % 5.207(5)
& $\phm2.33\phe$ % 2.329(2)
& $\phm0.29\phe$  % 0.2937(3)
& $\phm0.044\phe$ %  0.04419(5)
\\
$\si_{\LO,\mbox{\scriptsize $t/u$-int}}\ [\fb]$ 
& $-0.12\phe$ % $-0.1188(1)$
& $-0.091\phe$ % $-0.09070(6)$
& $-0.060\phe$ % $-0.05994(4)$
& $-0.016\phe$ % $-0.015537(7)$
& $-0.0034\phe$ % $-0.003426(1)$
\\
$\si_{\NLO,\mbox{\scriptsize $t/u$-int}}\ [\fb]$ 
& $-0.75\phe$ % $-0.753(1)$
& $\phm 0.17\phe$ % $ 0.1731(6)$
& $\phm 0.76\phe$ % $ 0.7574(8)$
& $\phm 0.089\phe$ % $ 0.08880(1)$
& $\phm 0.0044(1)$ % $ 0.00436(2)$
\\
\hline
\end{tabular}
}
\caption{As in \refta{ta:splits_nocuts}, but with VBF cuts applied.}
%\caption{$s$-channel and $t/u$-interference contributions to the 
%$\ppjjh$ cross section at LO and NLO, after imposing VBF cuts.}
\label{ta:splits_cuts}
\end{table}
The NLO result does not include the corrections due to photon-induced
processes, which cannot be split into the above subcontributions
respecting gauge invariance.

While $t/u$-interference terms, with or without VBF cuts, contribute
less than $1\%$ to the cross section, $s$-channel contributions are
clearly non-negligible when no cuts are used. At LO (NLO), for
$\MH=120\GeV$ they contribute $22\%$ ($27\%$) to the total cross
section, while for $\MH=200\GeV$ this contribution decreases to $9\%$
($10\%$).  For $\MH=700\GeV$ $s$-channel processes contribute less
than $1\%$ to the cross section, with and without VBF cuts. Thus, for
increasing Higgs-boson masses, the contribution from Higgs-strahlung
processes becomes less and less important compared to the contribution
from pure fusion processes.  When VBF cuts are used, both the
$s$-channel and $t/u$-channel-interference contributions are strongly
suppressed, yielding less than $0.6\%$ of the cross section for all
the studied Higgs-boson masses.  The comparably large NLO $s$-channel
contribution after VBF cuts originates from real gluon corrections
with up to three jets in the final state, because in contrast to LO
the two jets from the weak-boson decay, which tend to be aligned owing
to a boost, are not forced to be the two well-separated tagging jets.
We conclude that, applying typical experimental VBF cuts, the
contributions from $s$-channel diagrams and $t/u$-channel
interferences can be safely neglected.

\subsection{Leading-order b-quark contributions}
\label{se:b-contributions}

In this section we present the contributions arising at LO from
processes that include b-quarks in the initial and/or final states.
There are three types of contributions involving b~quarks. The first
type consists of $s$-channel diagrams with a
$\mathrm{b}\bar\mathrm{b}$ pair in the initial state, the second type
comprises $s$-channel diagrams with a $\mathrm{b}\bar\mathrm{b}$ pair
in the final state, and the third type involves $s$-channel diagrams
with a $\mathrm{b}\bar\mathrm{b}$ pair in both the initial and the
final state as well as all $t$- and $u$-channel diagrams where a b or
$\bar\mathrm{b}$~quark goes from the initial state to the final state.
In \refta{ta:bquarks_nocuts} we show, for different $\MH$ values, LO
cross-section results without b-quark contributions,
$\si_{\mathrm{LO,\; no\;b}}$, the results including only initial-state
b~quarks, $\si_{\LO,\;\bqin}$, the results including only final-state
b~quarks, $\si_{\LO,\;\bqout}$, and including both initial- and
final-state b~quarks, $\si_{\LO,\;\bqall}$.
\begin{table}
\def\phm{\phantom{-}}
\def\phn{\phantom{0}}
\centerline{
\begin{tabular}{|c|c|c|c|c|c|}
\hline
$\MH\ [\GeV]$ & 120 & 150 & 200 & 400& 700\\
\hline
$\si_{\mbox{\scriptsize LO, no b}}\ [\fb]$ 
& 5943(1)
& 4331(1)
& 2855.2(6)
&  900.7(1)
&  270.60(4)
\\
$\si_{\mbox{\scriptsize LO, b-in}}\ [\fb]$ 
& 5951(1)
& 4334(1)
& 2856.3(6)
&  900.7(1)
&  270.60(4) 
\\
$\de_{\mbox{\scriptsize b-in}}\ [\%]$ 
& 0.13(2)
& 0.07(2)
& 0.04(2)
& 0.01(2)
& 0.00(2)
\\
$\si_{\mbox{\scriptsize LO, b-out}}\ [\fb]$ 
& 6054(1)
& 4386(1)
& 2876.7(6)
&  902.4(1)
&  270.77(4)
\\
$\de_{\mbox{\scriptsize b-out}}\ [\%]$ 
& 1.87(2)
& 1.27(2)
& 0.75(2)
& 0.19(2)
& 0.06(2)
\\
$\si_{\mbox{\scriptsize LO, b-in/out}}\ [\fb]$ 
& 6203(1)
& 4495(1)
& 2945.7(6)
&  919.5(2)
&  274.49(4)
\\
$\de_{\mbox{\scriptsize b-in/out}}\ [\%]$ 
& 4.37(2)
& 3.79(2)
& 3.17(2)
& 2.09(2)
& 1.44(2)
\\
\hline
\end{tabular}
}
\caption{LO cross section for $\ppjjh$ with and without initial- and/or 
final-state b~quarks, without any cuts.}
\label{ta:bquarks_nocuts}
%\end{table}
%
%\begin{table}
\vspace{2em}
\def\phm{\phantom{-}}
\def\phn{\phantom{0}}
\centerline{
\begin{tabular}{|c|c|c|c|c|c|}
\hline
$\MH\ [\GeV]$ & 120 & 150 & 200 & 400 & 700\\
\hline
$\si_{\mbox{\scriptsize LO, no b}}\ [\fb]$ 
& 1876.1(5) 
& 1589.8(4)
& 1221.1(3)
&  487.32(9)
&  160.66(2)
\\
$\si_{\mbox{\scriptsize LO, b-in}}\ [\fb]$ 
& 1876.1(5) 
& 1589.8(4)
& 1221.1(3)
&  487.32(9)
&  160.66(2)
\\
$\de_{\mbox{\scriptsize b-in}}\ [\%]$ 
& 0.00(3)
& 0.00(2)
& 0.00(2)
& 0.00(2)
& 0.00(1)
\\
$\si_{\mbox{\scriptsize LO, b-out}}\ [\fb]$ 
& 1876.1(5) 
& 1589.8(4)
& 1221.1(3)
&  487.32(9)
&  160.66(2)
\\
$\de_{\mbox{\scriptsize b-out}}\ [\%]$ 
& 0.00(3)
& 0.00(2)
& 0.00(2)
& 0.00(2)
& 0.00(1)
\\
$\si_{\mbox{\scriptsize LO, b-in/out}}\ [\fb]$ 
& 1918.5(5) % 
& 1624.5(4) % 
& 1246.3(3) % 
&  495.55(9)
&  162.75(2) % 
\\
$\de_{\mbox{\scriptsize b-in/out}}\ [\%]$ 
& 2.26(3)
& 2.18(2)
& 2.06(2)
& 1.69(2)
& 1.30(1) 
\\
\hline
\end{tabular}
}
\caption{As in \refta{ta:bquarks_nocuts}, but with VBF cuts applied.}
%\caption{$\ppjjh$ LO cross section with and without initial- and/or
%final-state b~quarks and VBF cuts.}
\label{ta:bquarks_cuts}
\end{table}
The relative contributions arising from these subprocesses,
$\de_{\bqin}$, $\de_{\bqout}$, and $\de_{\bqall}$, are also shown. In \refta{ta:bquarks_cuts}
we present LO results including VBF cuts.

For low Higgs-boson masses and no cuts, final-state b~quarks increase
the LO cross section by up to $2\%$. The increase due to initial-state
b~quarks is one per mille or less, being strongly suppressed due to
the two bottom densities involved ($\mathrm{b}\bar\mathrm{b}$-annihilation
processes). Including both initial and final-state b~quarks increases
the total cross section by up to $4\%$, a contribution that is similar
in absolute value to the total EW correction, but opposite in sign.
The contributions from final-state and/or initial-state b~quarks
decrease with increasing Higgs-boson mass.  For $\MH=700\GeV$, b-quark
contributions from either final state or initial state become
negligible, while simultaneous initial- and final-state b-quark
corrections decrease to $1.4\%$.  When VBF cuts are imposed, b-quark
contributions become less important. This is particularly noticeable
in contributions arising from processes with final-state but no
initial-state b~quarks and vice versa, Higgs-strahlung processes of
the form $q \bar q \to \Pb \bar \Pb H$ and $\Pb \bar \Pb \to q \bar q
H$. These are $s$-channel processes and, as already shown in
\refse{se:further-split}, this type of contributions are strongly
suppressed by the VBF cuts.

\subsection{Scale dependence}

\begin{figure}
\vspace{2em}
%\framebox{
\includegraphics[bb= 85 440 285 660, scale=1]{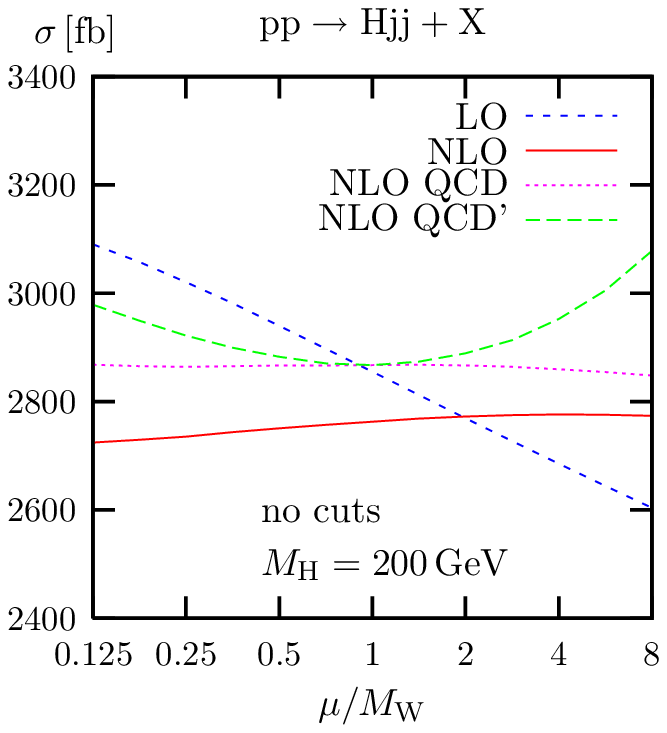}
%}
\quad
%\framebox{
\includegraphics[bb= 85 440 285 660, scale=1]{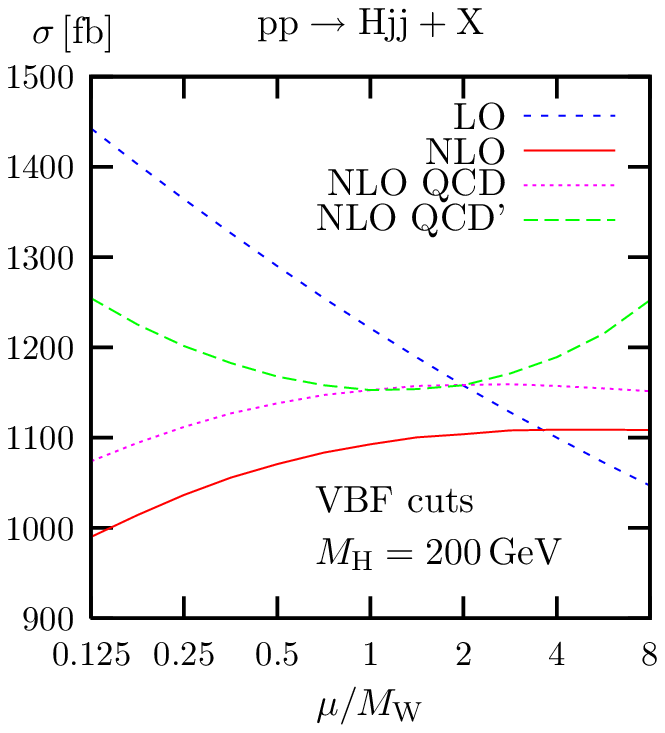}
%}
\caption{Scale dependence of LO and NLO cross section with QCD or QCD+EW
  corrections for $\MH=200\GeV$ without cuts (left)
  and with VBF cuts (right).}
\label{fi:mudep_200}
%\end{figure}
%%
%\begin{figure}
\vspace{2em}
%\framebox{
\includegraphics[bb= 85 440 285 660, scale=1]{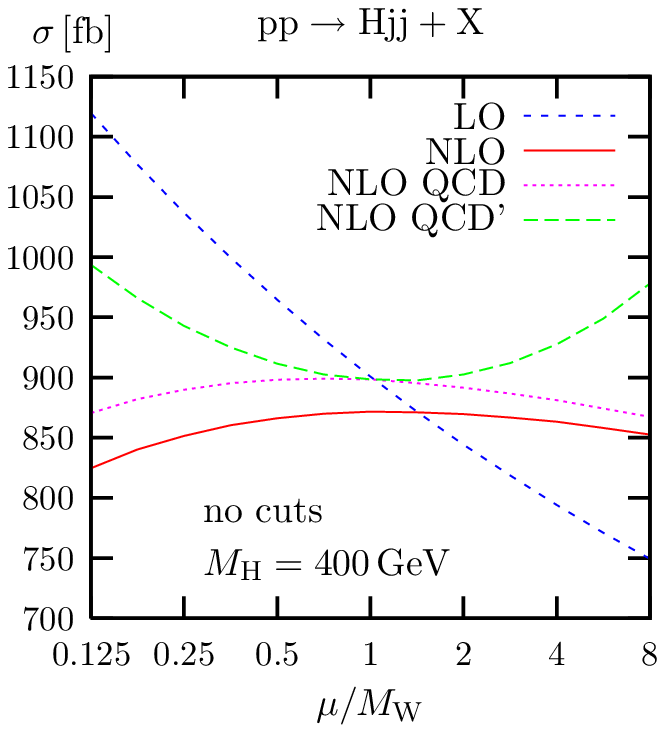}
%}
\quad
%\framebox{
\includegraphics[bb= 85 440 285 660, scale=1]{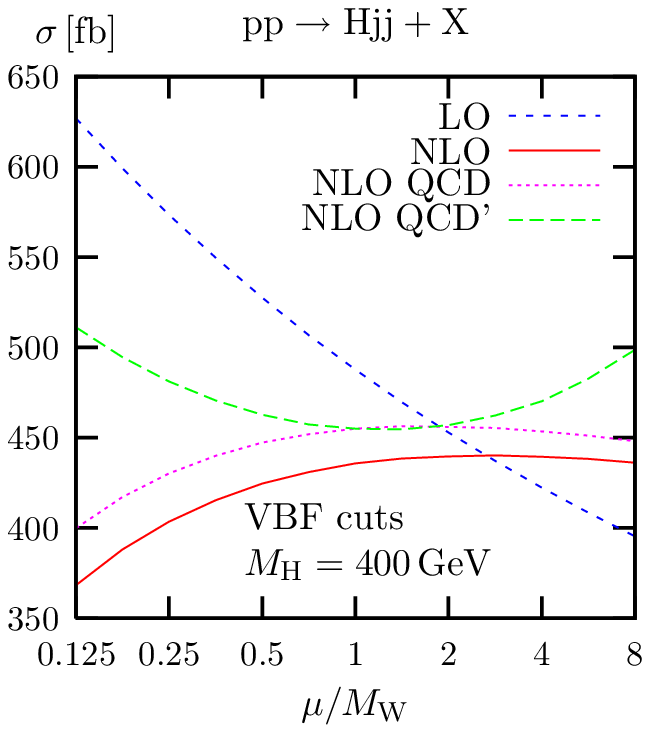}
%}
\caption{Scale dependence of LO and NLO cross section with QCD or QCD+EW
  corrections for $\MH=400\GeV$ without cuts (left)
  and with VBF cuts (right).}
\label{fi:mudep_400}
\end{figure}

In \reffis{fi:mudep_200} and \ref{fi:mudep_400} we show the dependence
of the total cross section on the factorization and renormalization
scale for $\MH=200\GeV$ and $\MH=400\GeV$, respectively.
We relate the factorization scale $\muF$, which applies to both QCD
and QED contributions, and the renormalization scale $\muR$ to the
W-boson mass as
\beq
\mu=\muF=\xiF\MW,\qquad
\muR=\xiR\MW,
\eeq
and vary $\xiF$ and $\xiR$ between $1/8$ and $8$.  We study the scale
dependence of the LO cross section, of the QCD-corrected NLO cross
section, and of the complete NLO cross section including both QCD and
EW corrections for $\xiR=\xiF$.  In addition we depict the
QCD-corrected NLO cross section for the setup where $\xiR=1/\xiF$ (NLO
QCD').
%For $\MH=120\GeV$, varying the scale up and down by a factor of 2 (8)
%changes the cross section by $\pm5\%$ ($\pm14\%$) in LO and by
%$\pm1\%$ ($\pm9\%$) in NLO for the set-up with VBF cuts.  Without
%cuts, the scale uncertainty is at the level of $\pm2\%$ ($\pm8\%$) in
%NLO, while it is accidentially small in LO for this specific
%Higgs-boson mass.  
For $\MH=200\GeV$, varying the scale up and down by a factor of 2 (8)
changes the cross section by $\pm3\%$ ($\pm9\%$) in LO and by $\pm1\%$
($-2\%/{+9\%}$) in NLO for the set-up without cuts. With VBF cuts, the
scale uncertainty amounts to $\pm6\%$ ($\pm18\%$) in LO and $\pm2\%$
($\pm11\%$) in NLO.  For $\MH=400\GeV$, the scale uncertainty is
reduced from $\pm7\%$ ($\pm24\%$) in LO to $\pm1\%$ ($\pm11\%$) in NLO
for the cross section without cuts, and from $\pm8\%$ ($\pm29\%$) in
LO to $\pm3\%$ ($\pm15\%$) in NLO for the cross section with VBF cuts.
For $\MH=400\GeV$, it is clearly seen from the results that $\MW$ is a
more appropriate scale choice than $\MH$. For this reason we have
chosen $\MW$ as default scale in this paper, while we used $\MH$ in
\citere{Ciccolini:2007jr}, where we only considered Higgs-boson masses
comparable to $\MW$.
%NLO cuts 200: 1.4, -9.4,7.1.10.6
%NLO cuts 400: -15.49, 0.09,10.1,11.7
%NLO nocuts 400: -5.4,-2.3,8.7,10.6

\subsection{Slicing cut dependence}

In the slicing approach (as \eg reviewed in \citere{Harris:2001sx}),
phase-space regions where real photon/gluon emission and
photon/gluon-induced processes contain soft or collinear singularities
are defined by the auxiliary cutoff parameters $\delta_\mathrm{s} ,\;
\delta_\mathrm{c}\ll 1.$ In real photon/gluon radiation processes, the
region
\beq
\la < k^0 < \delta_\mathrm{s} \frac{\sqrt{\hat s}}{2},
\eeq
where $k$ is the photon/gluon momentum, $\sqrt{\hat s}$ the partonic
centre-of-mass energy, and $\la$ an infinitesimal photon/gluon mass,
is treated in soft approximation.  The regions determined by
\beq
\label{eq:coll}
1-\cos(\theta_{\{\ga,\Pg\}q})  <  \delta_\mathrm{c}, \qquad
k^0  >  \delta_\mathrm{s}  \frac{\sqrt{\hat s}}{2},
\label{eq:slicingcuts}
\eeq
where $\theta_{\{\ga,\Pg\} q}$ is the angle between any quark $q$ and
the photon or gluon, are evaluated using collinear factorization. In
photon- or gluon-induced processes, singularities arise only in the
collinear region, i.e.\ a slicing cut on the angle between any
final-state quark $q$ and the initial-state photon or gluon is
sufficient to exclude the singularity from phase space.  Specifically,
we define this angular cut as in \refeq{eq:coll}.  The
collinear-splitting singularities are also treated using collinear
factorization.

\begin{figure}
%\framebox{
\includegraphics[bb= 85 440 285 660, scale=1]{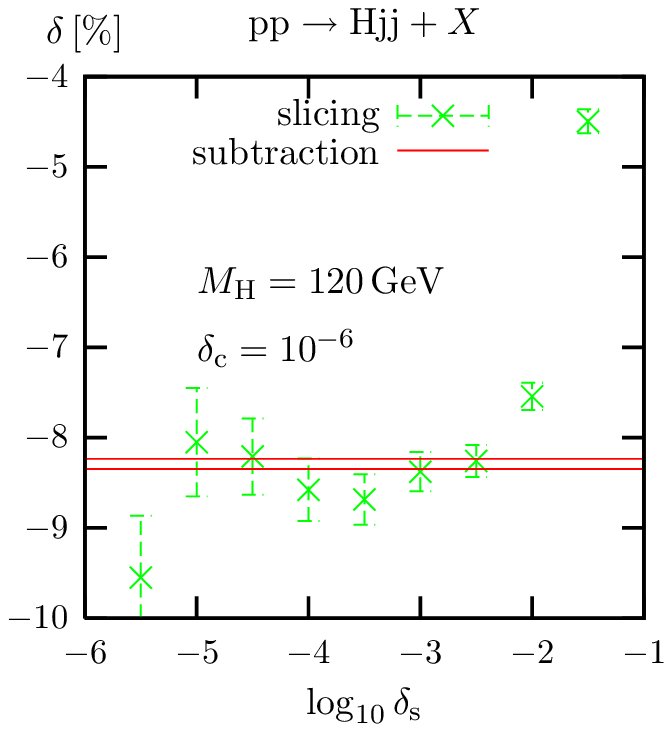}
%}
\quad
%\framebox{
\includegraphics[bb= 85 440 285 660, scale=1]{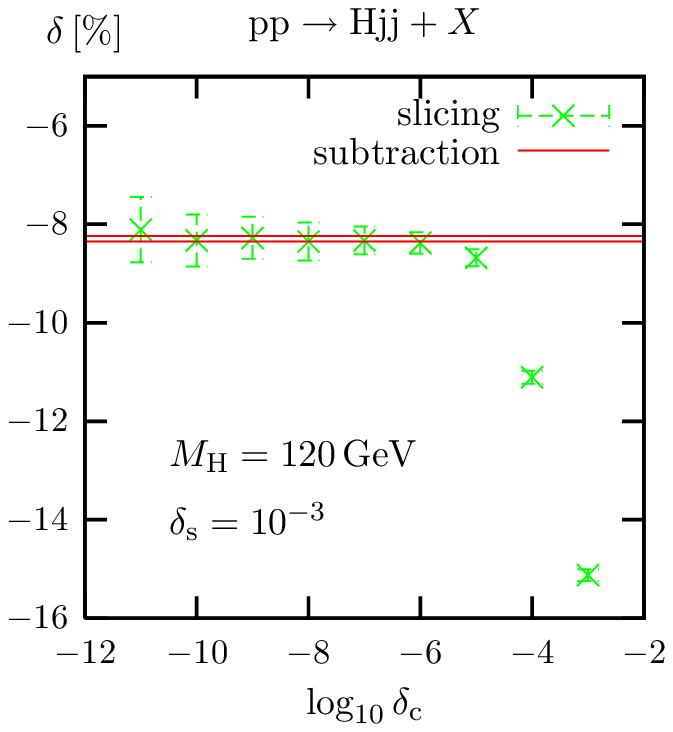}
%}
\caption{Dependence of the relative corrections to the total
  cross section with VBF cuts on the energy cutoff $\delta_\mathrm{s}$
  for $\delta_\mathrm{c}=10^{-6}$ (l.h.s.)\ and on the angular cutoff
  $\delta_\mathrm{c}$ for $\delta_\mathrm{s}=10^{-3}$ (r.h.s.)\ in the
  slicing approach for $\MH=120\GeV$. For comparison the corresponding
  result obtained with the dipole subtraction method (with 10 times
  less statistics) is shown as a $1\si$ band in the plots.}
\label{fi:sli}
\end{figure}

In the remaining phase space no regulators (photon/gluon and quark
masses) are used.  Therefore, the slicing result is correct up to
terms of ${\cal O}(\delta_\mathrm{s})$ and ${\cal
  O}(\delta_\mathrm{c})$.  In ~\reffi{fi:sli} we show the dependence
of the complete corrections to the cross section with VBF cuts on
$\delta_\mathrm{s}$ for fixed $\delta_\mathrm{c}=10^{-6}$ and the
dependence on $\delta_\mathrm{c}$ for fixed
$\delta_\mathrm{s}=10^{-3}$. The error bars reflect the uncertainty of
the Monte Carlo integration.
These results were obtained with $10^9$ events for the slicing method
and $10^8$ events for the subtraction method, using $\MH$ as
factorization and renormalization scale.  For decreasing auxiliary
parameters $\delta_\mathrm{s}$ and $\delta_\mathrm{c}$, the slicing
result reaches a plateau and becomes compatible with the subtraction
result. The integration error in the result obtained with the slicing
method increases for lower cut-off parameters.  On the other hand, the
subtraction results, for the same number of events, always show
smaller integration errors.

\subsection{Differential cross sections}

In this section we consider results for distributions involving
Higgs-boson and tagging-jet observables.  We show results for
$\MH=120\GeV$ in the setup including VBF cuts. For each distribution
we plot the absolute predictions in LO and in NLO including QCD and EW
corrections. In addition, we show the relative corrections, both the
QCD and EW corrections separately, as well as their sum.

\begin{figure}
%\framebox{
\includegraphics[bb= 85 440 285 660, scale=1]{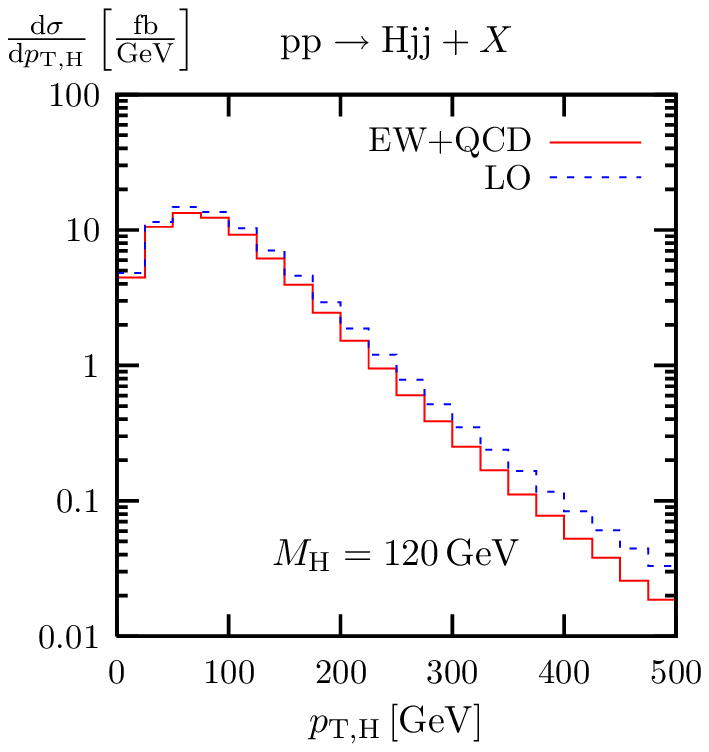}
%}
\quad
%\framebox{
\includegraphics[bb= 85 440 285 660, scale=1]{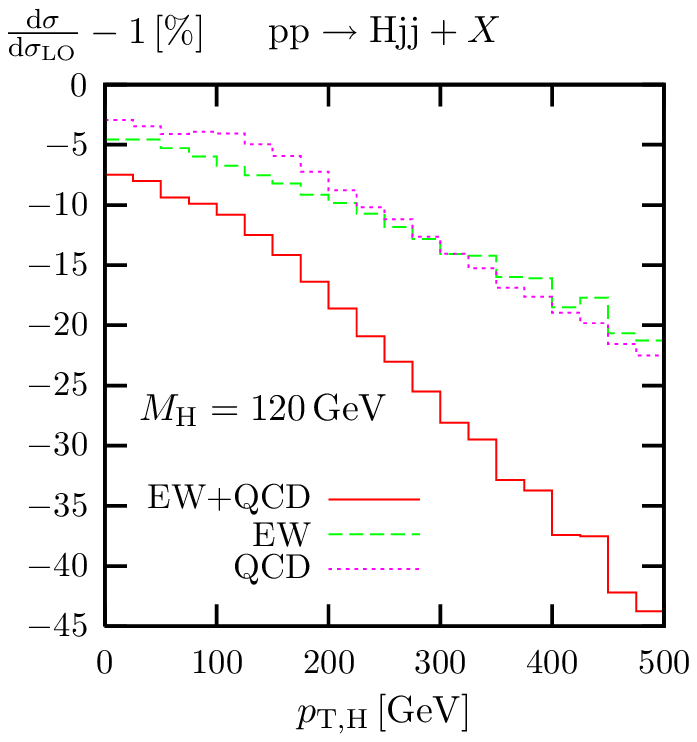}
%}
\caption{Distribution in the transverse momentum $p_{\rT,\PH}$ of the
  Higgs boson (left) and corresponding relative corrections (right)
  for $\MH=120\GeV$.}
\label{fi:pth}
%\end{figure}
%
%\begin{figure}
\vspace{2em}
\includegraphics[bb= 85 440 285 660, scale=1]{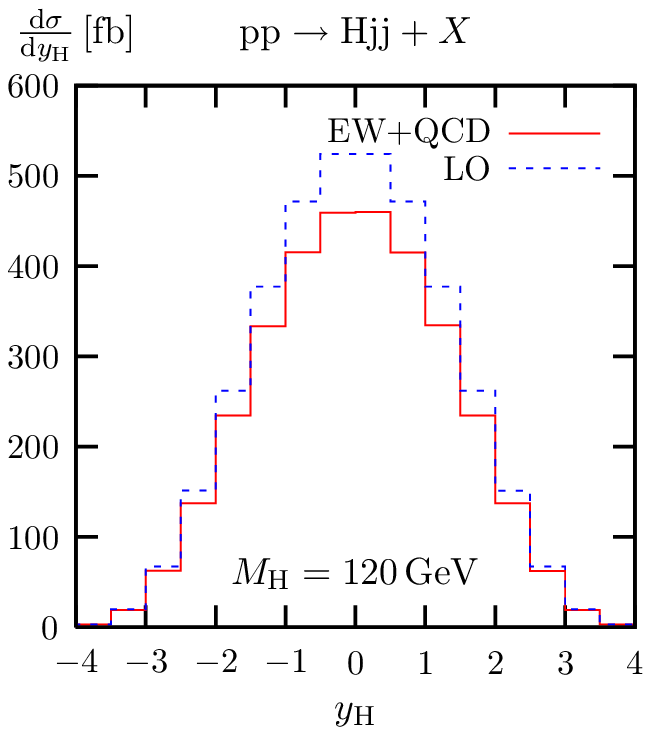}
\quad
\includegraphics[bb= 85 440 285 660, scale=1]{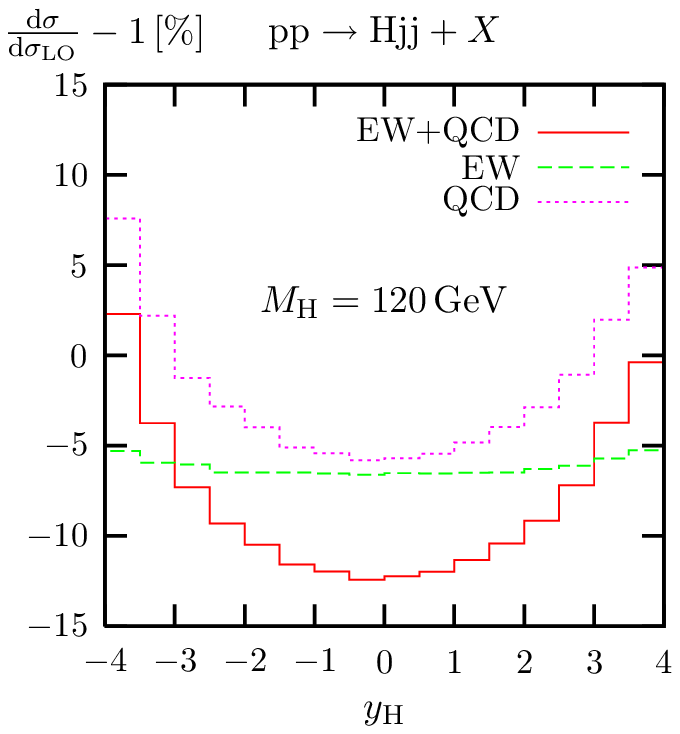}
\caption{Distribution in the rapidity $y_{\PH}$ of the Higgs boson
  (left) and corresponding relative corrections  (right) for $\MH=120\GeV$.}
\label{fi:yh}
\end{figure}

We first consider Higgs-boson observables and show the distribution
in the transverse momentum $p_{\rT,\PH}$ in \reffi{fi:pth}.
The differential cross section drops strongly with increasing
$p_{\rT,\PH}$, while both the relative EW and QCD corrections increase
in size and reach $-20\%$ for $p_{\rT,\PH}=500\GeV$.
It is interesting to note the differences between this result and the
same distribution in Higgs-boson production via gluon-fusion, as e.g.\ 
shown in \citere{Bozzi:2003jy}.  In weak-boson fusion, this
distribution is broader and peaks at a much larger value of
$p_{\rT,\PH}$.

The distribution in the rapidity $y_{\PH}$ of the Higgs boson is
presented in \reffi{fi:yh}.  While the relative EW corrections depend
only weakly on this variable, the QCD corrections show an increase for
large rapidities.  Total corrections decrease the differential cross
section by more than $10\%$ in the central region, inducing an
important change in the shape of this distribution.

\begin{figure}
  \includegraphics[bb= 85 440 285 660, scale=1]{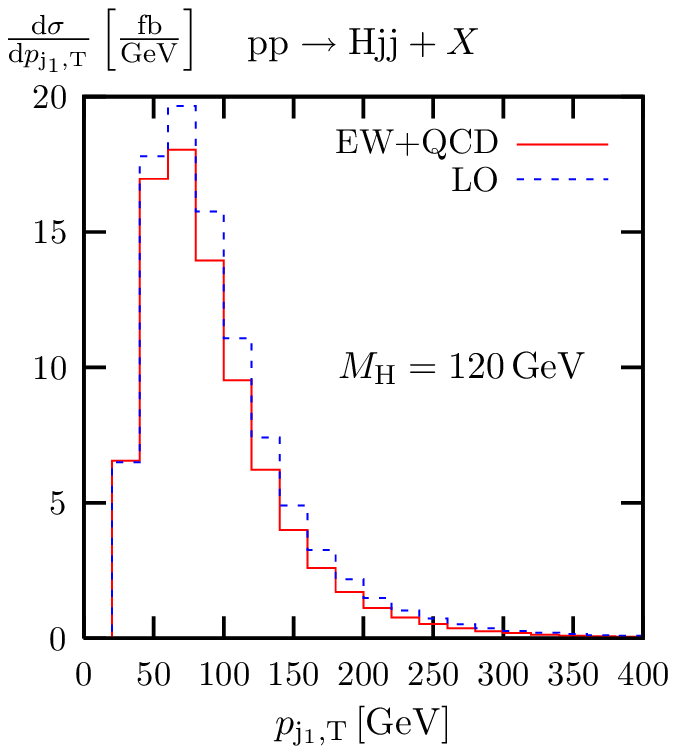} \quad
  \includegraphics[bb= 85 440 285 660, scale=1]{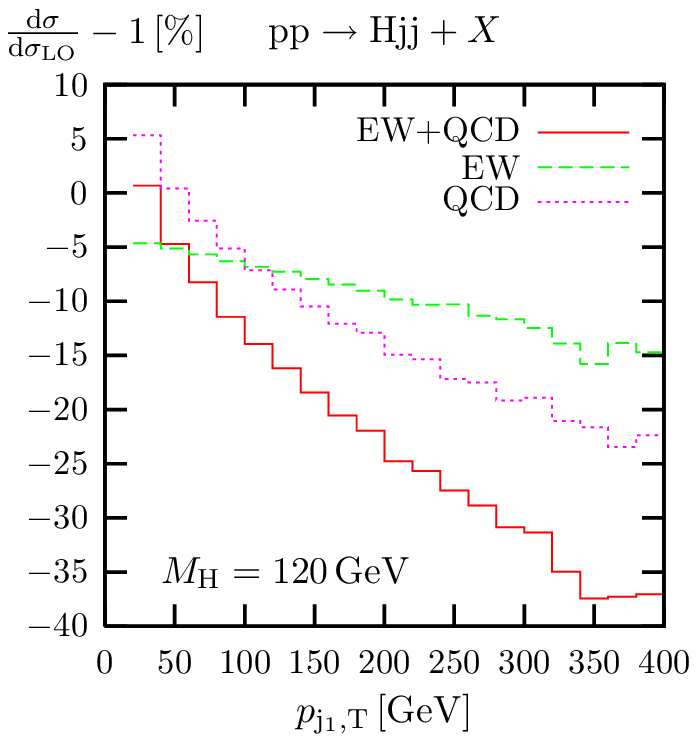}
\caption{Distribution in the transverse momentum
  $p_{\mathrm{j_1},\rT}$ of the harder tagging jet (left) and
  corresponding relative corrections (right) for $\MH=120\GeV$.}
\label{fi:pt1}
%\end{figure}
%\begin{figure}
\vspace{2em}
\includegraphics[bb= 85 440 285 660, scale=1]{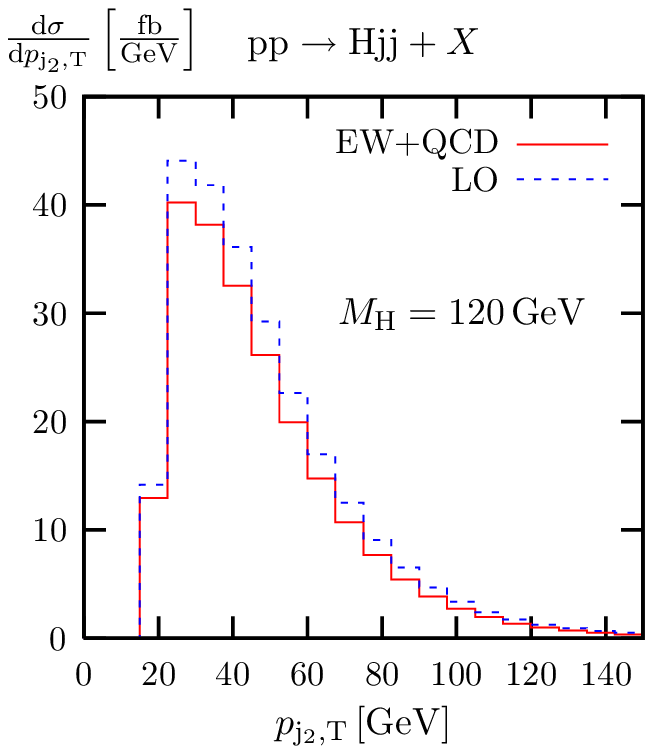}
\quad
\includegraphics[bb= 85 440 285 660, scale=1]{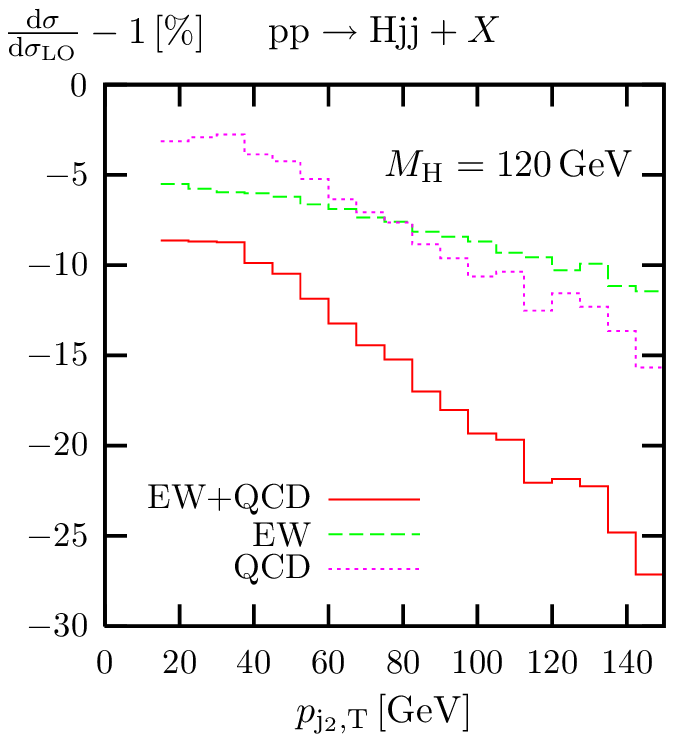}
\caption{Distribution in the transverse momentum
  $p_{\mathrm{j}_2,\rT}$ of the  softer tagging jet (left) and
  corresponding relative corrections (right) for $\MH=120\GeV$.} 
\label{fi:pt2}
\end{figure}

Figures \ref{fi:pt1} and \ref{fi:pt2} show the differential cross
section as function of the transverse momentum of the harder and
softer tagging jet, respectively. These distributions peak near or
below $p_\mathrm{j,T} \sim\MW$ and then drop strongly with increasing
jet transverse momentum. QCD and EW corrections become more and more
negative with increasing $p_\mathrm{j,T}$. For low transverse momentum
these corrections are at the level of 5\%, while for
$p_{\rj,T}=400\GeV$ and $150\GeV$ they add up to about $-38\%$ and
$-25\%$ for the harder and softer tagging jet, respectively. This
induces a substantial change in shape of these distributions.
%MH=200: Increasing the Higgs boson
%MH=200: mass to $\MH=200\GeV$ shifts both the QCD and EW corrections up, and the total
%MH=200: correction becomes positive for low $p_\mathrm{T}$ values.
%MH=200: 
%

\begin{figure}
\includegraphics[bb= 85 440 285 660, scale=1]{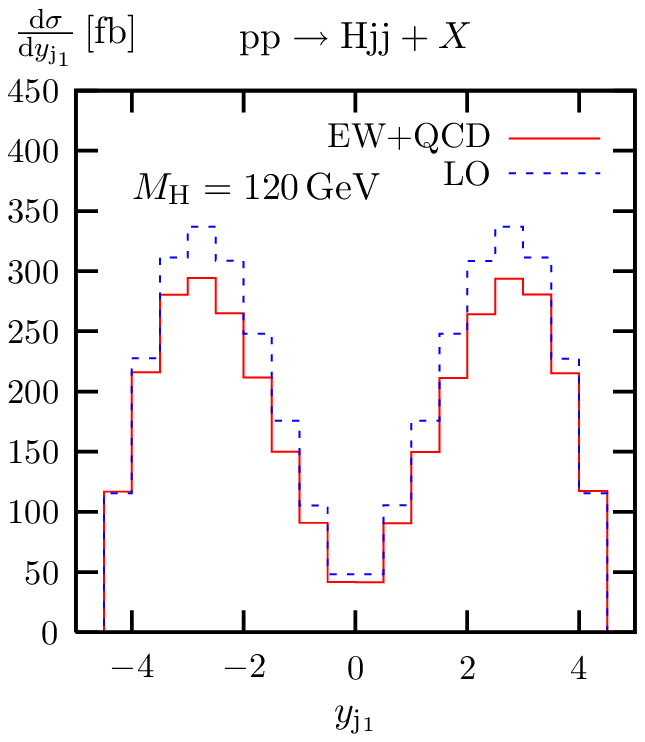}
\quad
\includegraphics[bb= 85 440 285 660, scale=1]{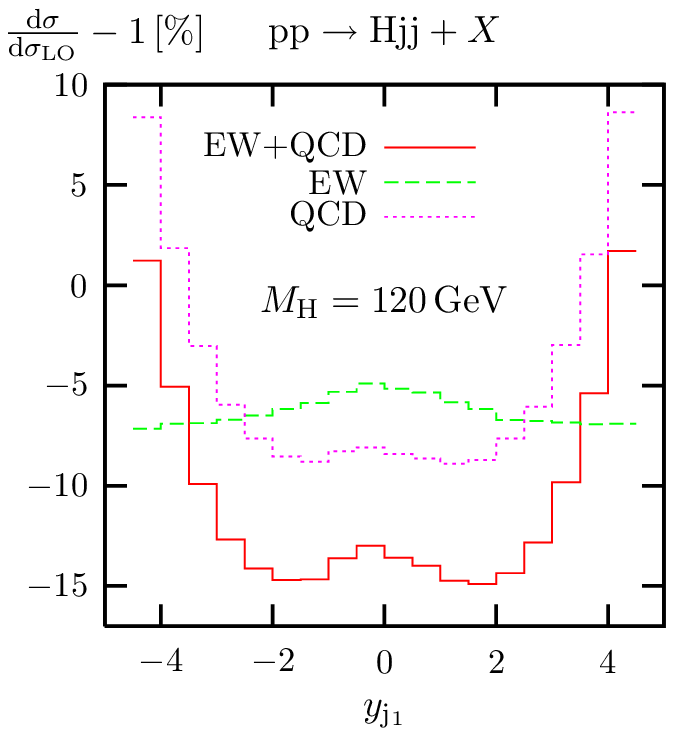}
\caption{Distribution in the rapidity $y_{\mathrm{j_1}}$ of the
  harder tagging jet (left) and corresponding relative corrections
  (right) for $\MH=120\GeV$.}
\label{fi:y1}
%\end{figure}
%\begin{figure}
\vspace{2em}
\includegraphics[bb= 85 440 285 660, scale=1]{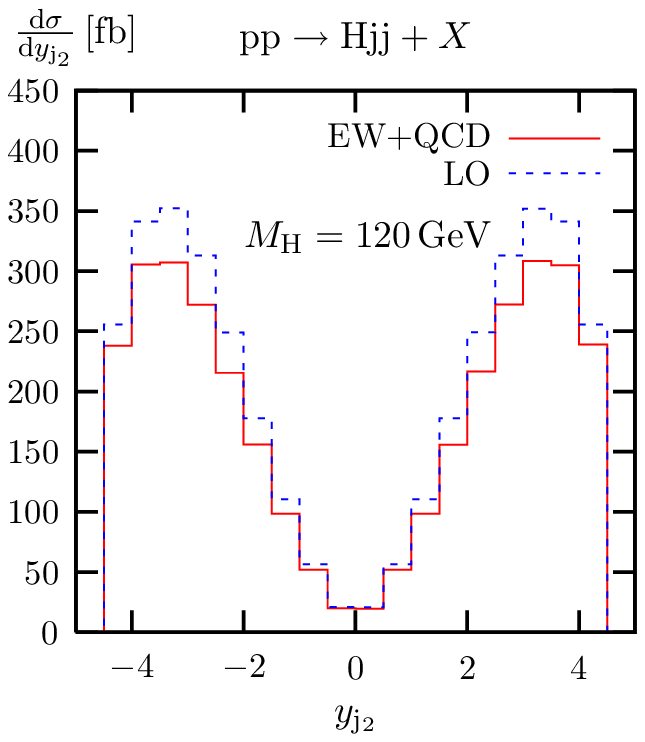}
\quad
\includegraphics[bb= 85 440 285 660, scale=1]{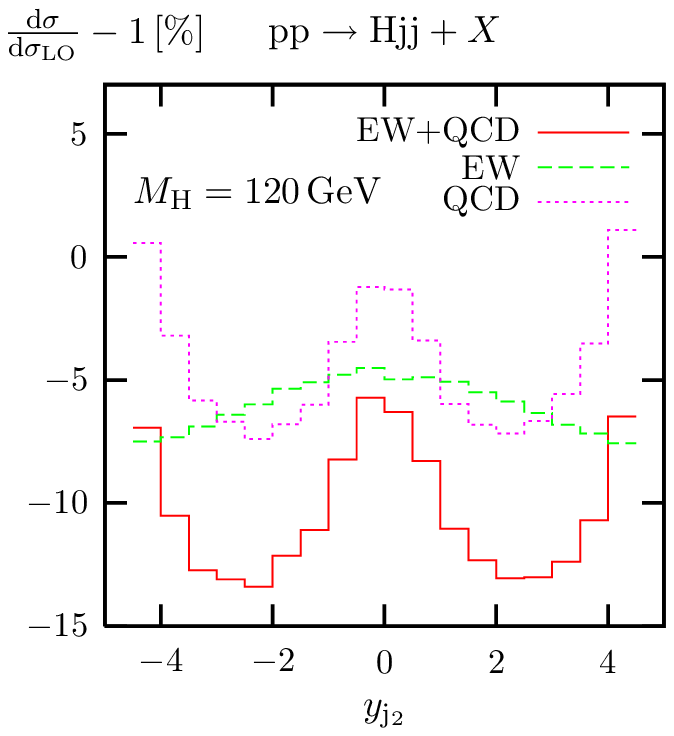}
\caption{Distribution in the rapidity $y_{\mathrm{j_2}}$ of the
  softer tagging jet (left) and corresponding relative corrections
  (right) for $\MH=120\GeV$.}
\label{fi:y2}
\end{figure}

In \reffis{fi:y1} and \ref{fi:y2}, we depict the distributions in the
rapidities of the harder and softer tagging jet, respectively.  It can
be clearly seen that the tagging jets are forward and backward
located.  The EW corrections vary between $-4\%$ and $-7\%$.  The QCD
corrections exhibit a strong dependence on the jet rapidities.  For
the harder tagging jet they are about $-8\%$ in the central region but
become positive for large rapidities, where they tend to compensate
the EW corrections. For the softer tagging jet the variation for large
rapidities is smaller, and the QCD corrections become small also near
$y_j=0$.  Shape changes due to the full corrections can reach $10\%$.
%MH=200: As in the case of jet transverse momenta, increasing the
%MH=200: Higgs-boson mass to $\MH=200\GeV$ shifts both the QCD and EW
%MH=200: corrections up. The total correction becomes positive for
%MH=200: large $y_j$ values, and, for the softer tagging jet, also 
%MH=200: for $y_j=0$.
%

\begin{figure}
  \includegraphics[bb= 85 440 285 660, scale=1]{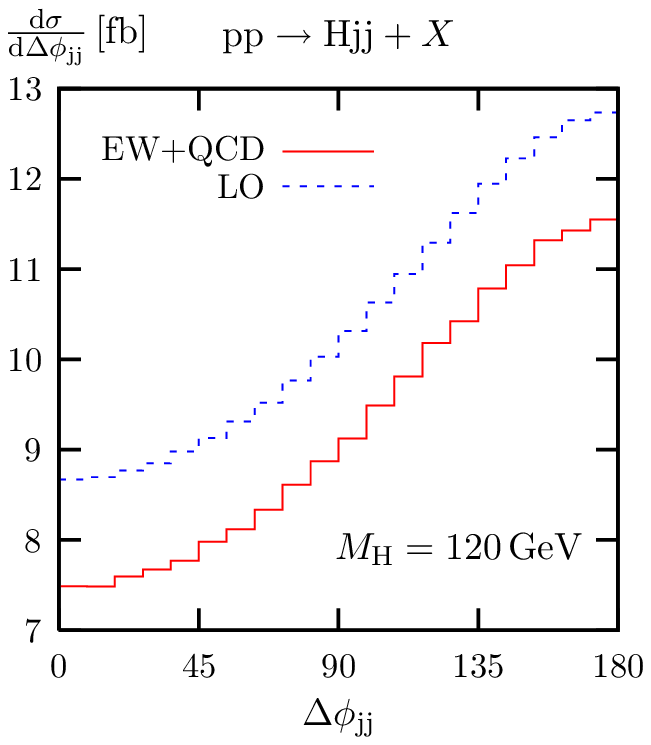}
  \quad \includegraphics[bb= 85 440 285 660,
  scale=1]{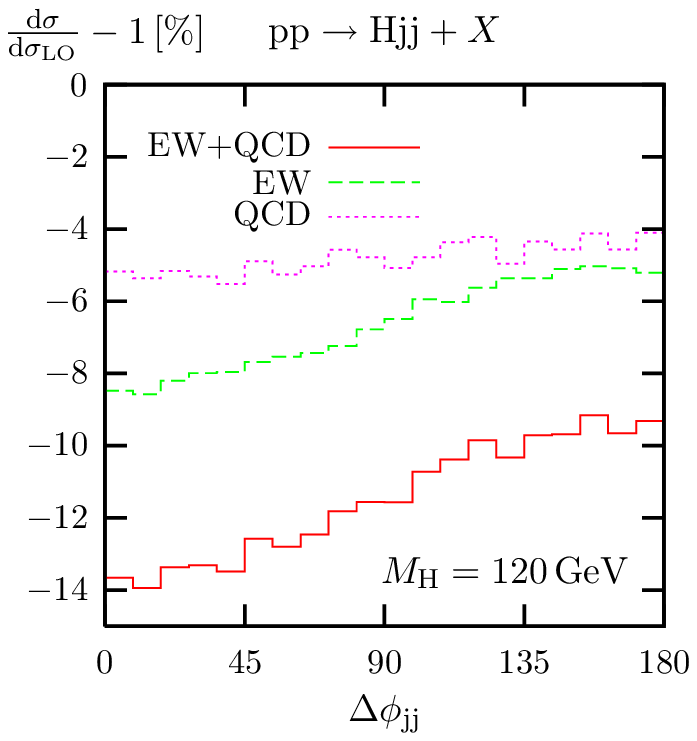}
\caption{Distribution in the azimuthal angle difference $\De
  \phi_{\mathrm{jj}}$ of the tagging jets (left) and corresponding
  relative corrections (right) for $\MH=120\GeV$.}
\label{fi:phi}
%\end{figure}
%
%\begin{figure}
\vspace{2em}
  \includegraphics[bb= 85 440 285 660, scale=1]{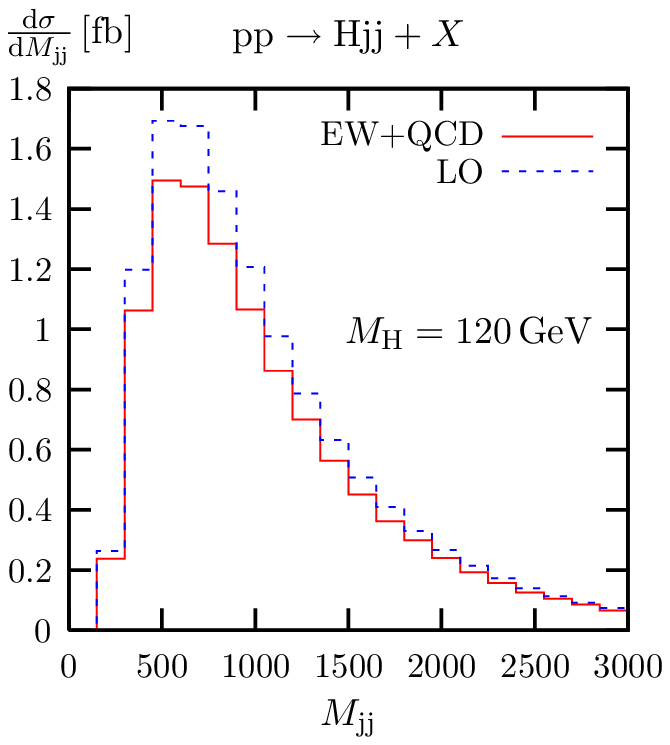}
  \quad \includegraphics[bb= 85 440 285 660,
  scale=1]{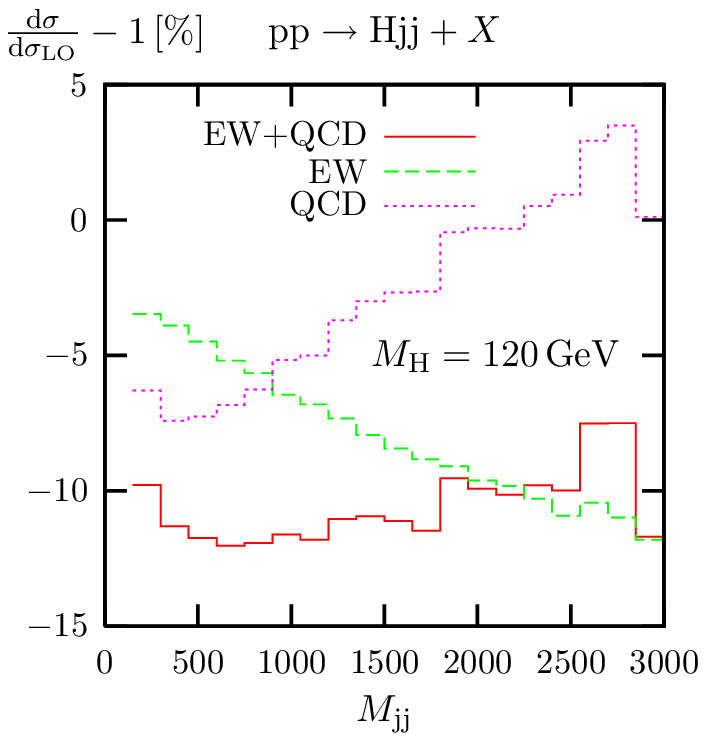}
\caption{Distribution in the tagging-jet-pair invariant mass
  $\mathrm{M}_{\mathrm{jj}}$ (left) and corresponding relative
  corrections (right) for $\MH=120\GeV$.} 
\label{fi:mjj}
\end{figure}

In \reffi{fi:phi} we present the distribution in the azimuthal angle
separation of the two tagging jets. This distribution is particularly
sensitive to non-standard contributions to the $\PH VV$ vertices
\cite{Figy:2004pt}.  As expected for VBF processes, there is a large
azimuthal angle separation between the two tagging jets.  While QCD
corrections are almost flat in this variable, the QCD+EW corrections
exhibit a dependence on $\De \phi_{\mathrm{jj}}$ on the level of 4\%.
%MH=200: For $\MH=200\GeV$ the total correction becomes almost $0$ for $\De \phi_{\mathrm{jj}}=180^\circ$.
%%

Finally, in \reffi{fi:mjj} we show the distribution in the
tagging-jet-pair invariant mass $M_{\mathrm{jj}}$.  Tagging jets
identified in EW processes have typically larger jet-pair invariant
masses than the ones identified in QCD processes.  Consequently,
$M_{\mathrm{jj}}$ can be used to further suppress QCD backgrounds, as
e.g.\ pointed out in \citere{Asai:2004ws}. This distribution peaks at
approximately $M_{\mathrm{jj}}=500\GeV$ and is strongly suppressed for
higher invariant-mass values.  The EW corrections decrease with
increasing $M_{\mathrm{jj}}$ and compensate the increasing QCD
corrections for large invariant masses. The total correction is of the
order of $-10\%$.
%MH=200: Increasing the Higgs-boson mass shifts the EW corrections, while the QCD corrections increase nontrivially.
%%

\subsection{Comparison with related NLO QCD calculations}
\newcommand{\offshell}{\mbox{\tiny off-shell}}
\newcommand{\vbfnlo}{\mbox{\tiny {\tt VBFNLO}}}
\newcommand{\vvtoh}{\mbox{\tiny {\tt VV2H}}}
\label{se:VBFcomparison}

In this section we compare our results to those obtained with the software 
packages {\tt VV2H} by M.~Spira \cite{vv2h:Spira} and {\tt VBFNLO} 
by D.~Zeppenfeld et al.\cite{vbfnlo:Zeppenfeld}%
\footnote{For this comparison we have employed the {\tt VV2H} version
  dated July 23 2007, and {\tt VBFNLO-v.1.0}.}.  These programs allow
to calculate the LO and NLO-QCD-corrected cross sections for
Higgs-boson production via VBF at hadron colliders.  It is important
to note that $s$-channel contributions and $t/u$-channel-interference
contributions are not taken into account in these calculations.  In
particular, only the $\mathcal{O}(\alpha_{\mathrm{s}})$ corrections
that correspond to our class (a) of QCD contributions (see
\refse{se:QCDcorr}) are included.  In order to allow for a tuned
comparison, we here use only four quark flavours for the external
partons, i.e.\ we have switched off the effect of initial- and
final-state b~quarks in the calculations. We compare the results of
{\tt VV2H} and {\tt VBFNLO} with LO and NLO-QCD-corrected results of
our code with $s$-channel contributions and $t/u$-channel-interference
contributions switched off, $\si^{\tuned}$. In addition, we give the
results of our code, $\si^{\best}$, with these contributions and all
interferences switched on and including all EW corrections apart from
photon-induced processes. We use CTEQ6 parton distributions
\cite{Pumplin:2002vw} and our default set of input parameters.
%Throughout this section we compare LO and NLO-QCD-corrected results
%for some Higgs masses between $120$ and $700\GeV$ using our default
%set of input parameters together with the CTEQ6 parton distributions
%\cite{Pumplin:2002vw}.

We first compare our results to those obtained with {\tt VV2H}, which
implements the formulae presented in \citere{Han:1992hr}.  As it is
not possible to include phase-space cuts in {\tt VV2H}, we have only
compared total cross sections.  The results of this comparison can be
found in \refta{ta:comp_vv2h}.  We observe that the LO cross sections
agree within $0.05\%$ and the NLO corrected results within $0.2\%$, a
difference which is of the order of the statistical error.
\begin{table}
\def\phm{\phantom{-}}
\def\phn{\phantom{0}}
\def\phpn{\phantom{.0}}
\centerline{
\begin{tabular}{|c|c|c|c|c|c|c|}
\hline
$\MH\ [\GeV]$                 & 120       & 150       & 170        & 200       & 400       & 700        \\ \hline
$\si^{\tuned}_{\LO}\ [\fb]$            & 4226.3(6) & 3357.8(5) & 2910.7(4)  & 2381.6(3) & 817.6(1)  & 257.49(4)  \\
$\si^{\vvtoh}_{\LO}\ [\fb]$   & 4226.2(4) & 3357.3(3) & 2910.2(3)  & 2380.4(2) & 817.33(8) & 257.40(3)  \\
$\si^{\best}_{\LO}\ [\fb]$    & 5404.8(9) & 3933.7(6) & 3290.4(5)  & 2597.9(4) & 834.5(1)  & 259.26(4)  \\\hline
$\si^{\tuned}_{\NLO}\ [\fb]$           & 4424(4)   & 3520(3)   & 3052(3)    & 2505(2)   & 858.4(7)  & 268.2(2)   \\
$\si^{\vvtoh}_{\NLO}\ [\fb]$  & 4415(1)   & 3519.7(8) & 3055.8(7)  & 2503.4(6) & 858.8(2)  & 268.03(6)  \\
$\si^{\best}_{\NLO}\ [\fb]$   & 5694(4)   & 4063(3)   & 3400(3)    & 2666(2)   & 839.0(7)  & 285.9(3)   \\\hline
\end{tabular}
}
\caption{Total cross section for $\ppjjh$ in LO and NLO 
calculated with our program, $\si_{\mathrm{LO/NLO}}$, and with 
{\tt VV2H}, $\si^{\vvtoh}_{\mathrm{LO/NLO}}$, for the setup defined 
in the text.}
\label{ta:comp_vv2h}
\end{table}
Our complete predictions $\si^{\best}$ differ from the results of VV2H
by up to 30\% for low Higgs-boson masses and by a few per cent for
high Higgs-boson masses.  The bulk of this big difference for small
$\MH$ values is due to the missing $s$-channel contributions in VV2H.

We now turn to {\tt VBFNLO}, which implements the results of
\citere{Figy:2003nv}. As explained there, {\tt VBFNLO} generates an
isotropic Higgs-boson decay into two massless ``leptons'' (which
represent $\tau^+\tau^-$ or $\gamma\gamma$ or $b\bar b$ final states),
and imposes a cut on the invariant mass of the Higgs boson.  In order
to be able to compare with this setup, we have implemented a
convolution with a Breit-Wigner distribution for the Higgs-boson in
one of our codes. When performing this convolution we can either
evaluate the matrix element for Higgs production for an on-shell Higgs
boson or for a Higgs boson with an invariant mass given by the
Breit-Wigner distribution. While the first variant is gauge invariant
and corresponds to a pole approximation, the second one, which is
implemented in VBFNLO, violates EW gauge invariance. Because of the
simple structure of the matrix element, this might not be a problem in
LO and if only QCD corrections are included.  Both variants neglect
contributions that do not involve a resonant Higgs boson, which is a
good approximation for small Higgs-boson masses, where the Higgs-boson
width is small, but not for large Higgs-boson masses, where the width
is large.  Using these two variants of our code, we have compared
cross sections without imposing any cuts on the decay products of the
Higgs boson and using a unit branching ratio.  To define the
integration region in the neighbourhood of the Higgs resonance, we
employ the value for the Higgs-boson width calculated by {\tt VBFNLO}.

The results for the cross section without cuts are compared in
\refta{ta:comp_nc}, while results including VBF cuts can be found in
\refta{ta:comp_vbfc}.  The relative difference between the results of
{\tt VBFNLO} and the variant of our code with off-shell matrix
elements, $\si^{\tuned}$, is below $0.04\%$ for the total LO cross
section and below $0.2\%$ for the NLO-QCD-corrected cross section,
both with and without VBF cuts.  This difference is of the order of
the statistical error. The difference between $\si^{\tuned}$ and the
variant with on-shell matrix elements, $\si^{\pole}$, is at the
per-mille level for Higgs-boson masses below $200\GeV$ but strongly
increases for a heavy Higgs boson. For $\MH=400\GeV$ and $700\GeV$ the
differences reach about $4\%$ and $30\%$, respectively, which
illustrates the order of uncertainty without a more sophisticated
treatment of off-shell effects of the Higgs boson including its decay.
The results for $\si^{\best}$ are obtained with on-shell matrix
elements only, since the off-shell matrix elements with EW corrections
become gauge dependent.  For small Higgs-boson masses and VBF cuts
applied these predictions differ from those of {\tt VBFNLO} by one per
mille or less in LO and by 6--8\%, the size of the EW corrections, in
NLO.  On the other hand, without cuts the big difference between
$\sigma^{\best}$ and the other predictions at small $\MH$ values is
again due to $s$-channel contributions.
\begin{table}
\def\phm{\phantom{-}}
\def\phn{\phantom{0}}
\def\phpn{\phantom{.0}}
\centerline{
\begin{tabular}{|c|c|c|c|c|c|c|}
\hline
$\MH\ [\GeV]$                 & 120       & 150       & 170          & 200          & 400          & 700       \\\hline
$\si^{\tuned}_{\LO}\ [\fb]$            & 4216.8(6) & 3350.0(5) & 2904.5(4)    & 2377.9(3)    & 824.8(1)     & 284.28(8) \\
$\si^{\vbfnlo}_{\LO}\ [\fb]$  & 4218.4(2) & 3351.1(2) & 2905.2(1)    & 2378.8(1)    & 825.06(5)    & 284.35(2) \\
$\si^{\pole}_{\LO}\ [\fb]$            & 4216.8(6) & 3349.7(5) & 2903.6(4)    & 2373.5(3)    & 786.1(1)     & 206.15(3) \\
$\si^{\best}_{\LO}\ [\fb]$    & 5394.0(9) & 3925.1(6) & 3282.5(5)    & 2590.5(4)    & 802.6(1)     & 207.75(3) \\\hline
$\si^{\tuned}_{\NLO}\ [\fb]$           & 4407(3)   & 3512(3)   & 3050(2)      & 2500(2)      & 865.5(6)     & 296.8(3)  \\
$\si^{\vbfnlo}_{\NLO}\ [\fb]$ & 4405.3(3) & 3512.0(2) & 3049.5(2)    & 2500.5(2)    & 866.32(7)    & 296.63(3) \\
$\si^{\pole}_{\NLO}\ [\fb]$           & 4409(3)   & 3511(3)   & 3043(4)      & 2494(2)      & 825.8(5)     & 214.7(1)  \\
$\si^{\best}_{\NLO}\ [\fb]$   & 5678(5)   & 4055(3)   & 3392(3)      & 2659(2)      & 808.1(5)     & 229.0(2)  \\\hline
\end{tabular}
}
\caption{Cross section for $\ppjjh$ in LO and NLO calculated with 
our program, $\si_{\mathrm{LO/NLO}}$, and with {\tt VBFNLO}, 
$\si^{\vbfnlo}_{\mathrm{LO/NLO}}$, without any cuts and
for the setup defined in the text.}
\label{ta:comp_nc}
%\end{table}
%
%\begin{table}
\vspace{2em}
\def\phm{\phantom{-}}
\def\phn{\phantom{0}}
\def\phnn{\phantom{00}}
\def\phpn{\phantom{.0}}
\def\phpnn{\phantom{.00}}
\centerline{
\begin{tabular}{|c|c|c|c|c|c|c|}
\hline
$\MH\ [\GeV]$                 & 120          & 150           & 170           & 200           & 400          & 700           \\\hline
$\si^{\tuned}_{\LO}\ [\fb]$            & 1683.2(3)    & 1430.6(2)     & 1287.3(2)     & 1104.6(1)     & 448.40(6)    & 159.44(3)     \\
$\si^{\vbfnlo}_{\LO}\ [\fb]$  & 1683.32(5)   & 1430.75(4)    & 1287.74(4)    & 1104.76(3)    & 448.41(1)    & 159.431(5)    \\
$\si^{\pole}_{\LO}\ [\fb]$            & 1682.6(3)    & 1430.4(2)     & 1287.5(2)     & 1103.6(1)     & 434.00(5)    & 123.13(1)     \\
$\si^{\best}_{\LO}\ [\fb]$    & 1682.9(3)    & 1429.6(2)     & 1287.0(2)     & 1103.2(1)     & 433.89(5)    & 123.11(1)     \\\hline
$\si^{\tuned}_{\NLO}\ [\fb]$           & 1726(1)      & 1459(2)       & 1307(1)       & 1118(1)       & 442.4(3)     & 155.0(2)      \\
$\si^{\vbfnlo}_{\NLO}\ [\fb]$ & 1725.3(2)    & 1458.9(1)     & 1308.8(1)     & 1117.6(1)     & 442.68(3)    & 154.71(1)     \\
$\si^{\pole}_{\NLO}\ [\fb]$           & 1724(2)      & 1460(1)       & 1310(2)       & 1118(1)       & 427.7(3)     & 118.0(1)      \\
$\si^{\best}_{\NLO}\ [\fb]$   & 1595(2)      & 1351(2)       & 1228(1)       & 1045(1)       & 403.2(3)     & 124.82(9)     \\\hline
\end{tabular}
}
\caption{As in \refta{ta:comp_nc}, but with VBF cuts applied.}
%\caption{Cross section for $\ppjjh$ in LO and NLO calculated with our 
%program, $\si_{\mathrm{LO/NLO}}$, and with {\tt VBFNLO}, 
%$\si^{\vbfnlo}_{\mathrm{LO/NLO}}$, for the setup defined in the text, 
%using VBF cuts.}
\label{ta:comp_vbfc}
\end{table}

\section{Conclusions}
\label{se:concl}

Higgs-boson production via weak-boson fusion is one of the most
important processes in the search for and the study of a Standard
Model-like Higgs boson at the LHC. In this paper we present the first
calculation of the NLO electroweak corrections for this process and we
extend previously existing approximate NLO QCD calculations by
including $s$-channel topologies (Higgs-strahlung processes) and all
interferences, both in LO and NLO.

We find that the electroweak corrections are of the order of
$5$--$10\%$, \ie as large as the NLO QCD corrections. Real corrections
induced by photons in the initial state increase LO results by roughly
$1\%$. More precisely, the electroweak corrections are approximately
$-5\%$ for Higgs masses below $200 \GeV$ and for larger $\MH$ values
steadily increase up to about $+7\%$ for $\MH=700 \GeV$. For this
Higgs-boson mass the leading two-loop effects in the heavy-Higgs
limit, which are included in our calculation, become as large as the
one-loop corrections. This signals the breakdown of perturbation
theory for large Higgs-boson masses. We suggest that the theoretical
uncertainty from missing higher-order corrections can be estimated by
the size of the leading two-loop heavy-Higgs effects in this domain.
Moreover, for $\MH\gsim400\GeV$ owing to the large Higgs-boson width
the on-shell approximation is not sufficient any more and a more
sophisticated treatment including off-shell effects of the Higgs boson
and its decay width is required.

We have implemented our calculation in a flexible Monte Carlo event
generator, and studied differential distribution in Higgs-boson and
tagging-jet observables. Specifically, we have presented results for
distributions in transverse momenta, in rapidities, in the azimuthal
angle difference of the tagging jets, and in the tagging-jet pair
invariant mass. We found that QCD and electroweak corrections do not
simply rescale differential distributions, but induce distortions at
the level of $10\%$.

Finally, we have compared our NLO QCD-corrected results with existing
calculations, which only take into account $t/u$-channel
squared-diagram contributions. Working in this approximation, which
renders the QCD corrections particularly simple, we found technical
agreement between our results and the existing calculations within
statistical integration errors. We also found that, when typical VBF
cuts are applied, our full NLO QCD results agree with the ones in the
$t/u$-channel approximation within fractions of a per cent.

With the complete knowledge of NLO QCD and electroweak corrections,
the theoretical uncertainty from missing higher-order effects should
be of the order of 1--2\% in total cross-section predictions for
Higgs-boson masses in the range 100--$200\GeV$. For distributions, the
uncertainty will be larger in suppressed phase-space regions. The
phenomenological error of the parton distributions contributes a
further $3.5\%$ to the uncertainty, as reported in
\citere{Figy:2003nv}. We thus conclude that the presented
state-of-the-art results match the required precision for predictions
at the LHC.

\section*{Acknowledgements}

We thank M.~Spira and D.~Zeppenfeld for useful discussions.  This work
is supported in part by the European Community's Marie-Curie Research
Training Network under contract MRTN-CT-2006-035505 ``Tools and
Precision Calculations for Physics Discoveries at Colliders''.
Finally, we thank the Galileo Galilei Institute for Theoretical
Physics in Florence for the hospitality and the INFN for partial
support during the completion of this work.

\end{document}